\documentclass[apj]{emulateapj}
\usepackage{graphicx}
\usepackage{subfigure}
\usepackage {natbib}

\slugcomment{Received DATE; accepted DATE}

\begin{document}

\title{The Kinematics of Multiple-Peaked Ly$\alpha$ Emission in Star-Forming 
Galaxies at $z\sim2-3$\altaffilmark{1}}

\shorttitle{MULTIPLE-PEAKED LY$\alpha$ EMISSION AT $Z\sim2-3$}
\shortauthors{KULAS ET AL.}

\author{\sc Kristin R. Kulas, Alice E. Shapley\altaffilmark{2}}
\affil{Department of Astronomy, University of California,
Los Angeles, 430 Portola Plaza, Los Angeles, CA 90095-}

\author{\sc Juna A. Kollmeier}
\affil{Observatories of the Carnegie Institution of Washington, 
813 Santa Barbara Street, Pasadena, CA 91101, USA}

\author{\sc Zheng Zheng}
\affil{Yale Center for Astronomy and Astrophysics,
Department of Physics,
Yale University,
P.O. Box 208121,
New Haven, CT 06520-8121}
\affil{Department of Physics and Astronomy, 
University of Utah, 
115 South 1400 East, 
Salt Lake City, UT 84112}

\author{\sc Charles C. Steidel}
\affil{California Institute of Technology,
MS 249-17,
Pasadena, CA 91125}

\and

\author{\sc Kevin N. Hainline}
\affil{Department of Astronomy, University of California,
Los Angeles, 430 Portola Plaza, Los Angeles, CA 90024}

\altaffiltext{1}{Based, in part, on data obtained at the W.M. Keck
Observatory, which is operated as a scientific partnership among the
California Institute of Technology, the University of California, and
NASA, and was made possible by the generous financial support of the W.M.
Keck Foundation.}

\altaffiltext{2}{David and Lucile Packard Fellow}

\begin{abstract}

We present new results on the Ly$\alpha$ emission-line kinematics of 18 $z\sim2-3$ 
star-forming galaxies with multiple-peaked Ly$\alpha$ profiles.  With our large 
spectroscopic database of UV-selected star-forming galaxies at these 
redshifts, we have determined that 
$\sim$30$\%$ of such objects with detectable Ly$\alpha$ 
emission display multiple-peaked emission profiles.  
These profiles provide additional constraints on 
the escape of Ly$\alpha$ photons due to the rich velocity structure in the emergent 
line.  Despite recent advances in modeling the escape of Ly$\alpha$ from 
star-forming galaxies at high redshifts, comparisons between models and data are 
often missing crucial observational information.  Using Keck II NIRSPEC spectra of 
H$\alpha$ ($z\sim2$) and [OIII]$\lambda$5007 ($z\sim3$), we have measured accurate 
systemic redshifts, rest-frame optical nebular velocity dispersions and 
emission-line fluxes for the objects in the sample.  In addition, rest-frame UV 
luminosities and colors provide estimates of star-formation rates (SFRs) and the 
degree of dust extinction.  In concert with the profile sub-structure, these 
measurements provide critical constraints on the geometry and kinematics of interstellar
gas in high-redshift galaxies.  Accurate systemic redshifts allow us to translate the 
multiple-peaked Ly$\alpha$ profiles into velocity space, revealing that the 
majority (11/18) display double-peaked emission straddling the velocity-field 
zeropoint with stronger red-side emission.  Interstellar absorption-line kinematics 
suggest the presence of large-scale outflows for the majority 
of objects in our sample, with an 
average measured interstellar absorption velocity offset of $\langle \Delta 
v_{abs} \rangle =-230$ km s$^{-1}$.  A comparison of the interstellar absorption 
kinematics for objects with multiple- and single-peaked Ly$\alpha$ profiles 
indicate that the multiple-peaked objects are characterized by significantly 
narrower absorption line widths. We compare our data with the predictions
of simple models for outflowing and infalling gas distributions around high-redshift galaxies. While
popular ``shell" models provide a qualitative match with many of
the observations of Ly$\alpha$ emission, we find that in detail
there are important discrepancies between the models and data,
as well as problems with applying the framework of an expanding thin shell of gas
to explain high-redshift galaxy spectra.
Our data highlight these inconsistencies, as well as illuminating
critical elements for success in future models of outflow and infall in high-redshift galaxies.

\end{abstract}


\section{Introduction}

The process described as ``feedback" is considered a crucial component in models of 
galaxy formation. Feedback commonly refers to large-scale outflows of mass, metals, 
energy, and momentum from galaxies, which therefore regulate the amount of gas 
available to form stars, as well as the thermodynamics and chemical enrichment of 
the surrounding intergalactic medium (IGM). The evidence for feedback in 
high-redshift star-forming galaxies comes in several forms, including blueshifts of 
hundreds of km s$^{-1}$ in interstellar absorption lines relative to galaxy 
systemic redshifts \citep{pettini01,shapley03,steidel10}, and the nature of the IGM 
environments of vigorously star-forming galaxies, in terms of the optical depth and 
kinematics of the surrounding H I and heavy elements 
\citep{adelberger03,adelberger05}. Yet, in spite of evidence for ubiquitous 
outflows at high redshift, estimates of fundamental physical quantities such as gas 
column densities and mass outflow rates in the superwinds have remained elusive.
While there is much observational evidence for outflows in high-redshift galaxies,
both analytic models and hydrodynamical simulations of these systems
suggest that, at the same cosmic epochs, they should
be rapidly accreting cold gas from the IGM, fueling their 
active rates of star formation \citep{birnboimdekel2003,keres2005,keres2009,dekel2009}.
Searching for observational signatures of the process of cold gas accretion
at high redshift remains an open challenge.

The Ly$\alpha$ feature is one of the most widely used probes of star formation in 
both the nearby and very distant universe. While Ly$\alpha$ photons are initially 
produced by recombining ionized gas in H II regions, resonant scattering through 
the interstellar medium (ISM) of galaxies can lead to extreme modulation of the 
intrinsic emission profile, in both frequency and spatial location. Absorption by 
dust can completely suppress Ly$\alpha$ emission, producing a strong absorption 
profile, even in galaxies with high rates of star formation. It is therefore 
difficult to determine the intrinsic properties of the gas giving rise to 
Ly$\alpha$ emission from the observed profile alone. Only recently has it become 
possible to make detailed theoretical predictions for the Ly$\alpha$ profiles 
emergent from complex systems similar to those observed at high redshift 
\citep{ahn02,zheng02,hansenoh06,verhamme06,verhamme08}. These new calculations 
\citep[e.g.,][]{verhamme06} employ a Monte Carlo approach to propagate a 
representative ensemble of Ly$\alpha$ photons through gas and dust of arbitrary 
spatial and velocity distribution outputting Ly$\alpha$ profiles as would be 
observed. By comparison with Ly$\alpha$ profiles in actual galaxy spectra, it is 
possible, in principle, to recover quantities such as the expansion/infall velocity of the 
outflow/inflow, the column density and velocity dispersion of absorbing gas, and the gas 
covering fraction. Therefore, modeling of observed Ly$\alpha$ emission profiles 
represents an independent method of probing the processes of feedback and accretion at high 
redshift.

The majority of Ly$\alpha$ emission profiles at high redshift fall in the category 
of single-peaked and asymmetric \citep{shapley03,tapken07}, which is a natural 
outcome of an expanding medium \citep{verhamme08}.  However, multiple-peaked 
Ly$\alpha$ profiles, seen in a fraction of star-forming UV-selected galaxies at 
$z\sim2-3$, offer a particularly detailed perspective on Ly$\alpha$ escape due to 
the rich structure of the emergent line.  In principle, the structure of the 
Ly$\alpha$ line encodes the velocity field and density distribution of the gas 
through which it has emerged.  For example, a symmetric double-peaked profile 
centered on the velocity-field zeropoint is a natural outcome of the radiative 
transfer of Ly$\alpha$ photons through a static medium \citep{osterbrock1962}.  
Recent advances in modeling the escape of the Ly$\alpha$ photons from star-forming 
galaxies at $z\sim2-3$ have isolated several features of the Ly$\alpha$ profile 
that may be expected for specific gas geometries and velocity fields 
\citep[e.g.][]{verhamme06,laursen2009a,laursen2009b,barnes2011}. 
Since Ly$\alpha$ is the most readily 
observed high-redshift emission line, there is enormous potential to better understand the 
structure of the early galaxies by interpreting Ly$\alpha$ line profiles.

Previous attempts to compare observed Ly$\alpha$ line morphologies to theoretical 
predictions have been missing crucial information, weakening the derived constraints 
on galaxy outflow and inflow properties.  
In particular, accurate systemic redshift measurements, 
nebular line widths, and intrinsic ionizing photon fluxes have been absent from 
most previous comparisons \citep[but see, e.g.,][]{yang2011,mclinden2011}. 
These three observables are critical for anchoring the 
velocity scale of the models, constraining the mass and thermal motions of the 
gas, and determining the overall normalization for a given model.  
In this paper, we use new results obtained from H$\alpha$ and 
[OIII]$\lambda$5007 emission lines to provide the critical missing constraints 
on the observed kinematics of star-forming galaxies at $z\sim2-3$ with 
multiple-peaked Ly$\alpha$ emission.  In addition, we have used our large database 
of Ly$\alpha$ emission lines in high-redshift objects to determine the 
$\it{frequency}$ of the multiple-peaked systems, providing a global perspective on 
the potential of using Ly$\alpha$ morphology to reveal $z\sim2-3$ gaseous 
structure.

Our method of target selection from the parent sample of UV-selected
galaxies at $z\sim2-3$ is explained in Section~\ref{sec:samp}, while
the observations and data reduction are described in Section \ref{sec:obsdata}. 
The Ly$\alpha$ velocity profiles are presented in Section 
\ref{sec:lyavel} with precise velocity-field zeropoints determined from our 
H$\alpha$ (or [OIII]$\lambda5007$) measurements.  In Section \ref{sec:physquant} 
the measured physical quantities for each system are reported.  Section 
\ref{sec:model} describes current Ly$\alpha$ radiative transfer models, in addition 
to a qualitative comparison between some simple models and our measured
Ly$\alpha$ velocity profiles.  
Finally, in Section \ref{sec:conclusions} we summarize our results and discuss how 
our measurements will be used in future work to accurately model the processes that 
give rise to the multiple-peaked Ly$\alpha$ profiles from star-forming galaxies at 
$z\sim2-3$.  We assume a flat $\Lambda$CDM cosmology with $\Omega_{m}=0.3$, 
$\Omega_{\Lambda}=0.7$, and $H_{o}=70$ km s$^{-1}$ Mpc$^{-1}$.

\section{Sample Selection}
\label{sec:samp}

\begin{figure*}
\begin{center}
\centerline{
%   \mbox{\includegraphics[scale =0.8]{images2_xfig.eps}}
   \mbox{\includegraphics[scale =0.80]{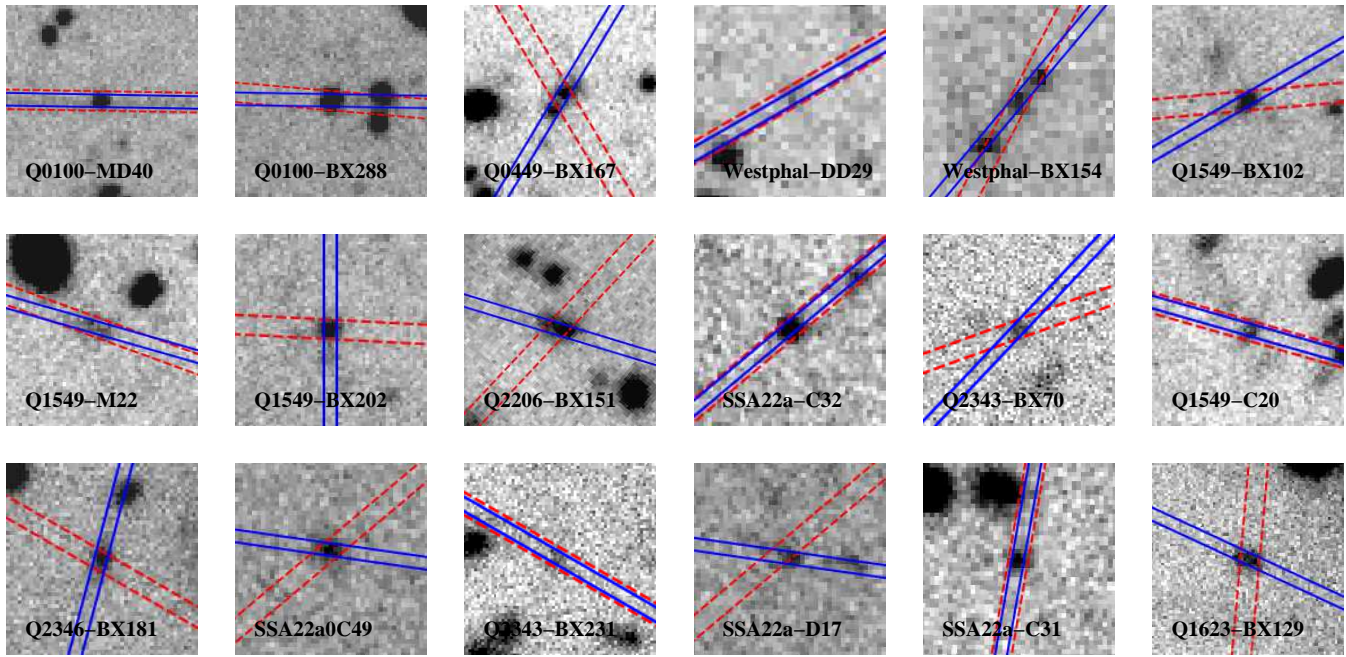}}
  }
\caption{$\cal R$-band images (rest-frame UV) for our sample of 18 objects.  In each image, the NIRSPEC long slit position is overlaid with blue, solid lines.  The LRIS slit is overlaid with red, dashed lines.  All images are $12^{\prime\prime}$$\times$$12^{\prime\prime}$, with north up and east to the left.  Typical seeing during these exposures was $0.^{\prime\prime}8-1.^{\prime\prime}0$.
\label{fig:spec1}}
\end{center}
\end{figure*}

Our target galaxies were drawn from the UV-selected sample described in 
\citet{steidel2003,steidel2004}.  These UV-selected surveys have yielded over 3,000 
spectroscopically confirmed star-forming galaxies at $z\sim1.5-3.5$.  The majority 
of the spectra cover the HI Ly$\alpha$ feature at 1216 \AA.  The observed Ly$\alpha$ 
profiles for UV-selected star-forming galaxies vary widely in strength, from damped 
absorption to strong emission.  Of the $\sim$1500 objects that show net emission, a 
fraction exhibit a multiple-peaked profile.  While this phenomenon has been 
previously reported \citep{tapken07,quider09} its frequency of occurrence has never 
been systematically analyzed.  We have used our database of star-forming galaxies to 
assess the frequency of multiple-peaked Ly$\alpha$ profiles in $z\sim2-3$ galaxies 
in order to determine whether such profiles represent rare outliers or are 
commonplace features of the galaxy population.

To identify the frequency of multiple-peaked profiles, we first separated the full 
spectroscopic sample into subsets of galaxies with and without detectable Ly$\alpha$ 
emission.  The $\sim$1500 galaxies with Ly$\alpha$ emission were then considered for 
further study.  All of the optical (rest-frame UV) spectra were obtained using the 
LRIS spectrograph on the Keck I telescope \citep{oke95}.  These spectra were taken 
with the 300 line mm$^{-1}$ grating blazed at 5000 \AA, or, following the
LRIS-B upgrade \citep{steidel2004}, with the 400 line mm$^{-1}$ or 
600 line mm$^{-1}$ grism blazed at 3400 \AA, and 4000 \AA, respectively.
The resolution of the spectra (taken through 1.$^{\prime\prime}$2 slits) establishes a 
lower limit in velocity space on the separation of multiple peaks that can be 
identified.  From this sample we were able to identify objects that have a minimum 
peak separation of 225 [370, 500] km s$^{-1}$ for the 600 [400,300]-line grism/grating.  We 
separated the spectra by grism/grating and analyzed each resolution group 
individually. Approximately 60$\%$ of the spectra were obtained with the 400- and 600-line grisms, while the rest of the spectra were taken using the 300-line grating.

We used a systematic algorithm to search for multiple peaks within the Ly$\alpha$ 
profile.  Several criteria were used to minimize contamination from noisy spectra 
which can yield a false indication of multiple peaks due to noise spikes.  The first 
criterion was a signal-to-noise (S/N) lower limit of 3 on the strongest flux peak height, 
which was implemented to establish a consistent threshold for characterizing 
Ly$\alpha$ profiles.  This criterion reduced our sample to $\sim$1000 spectra with 
163, 527, and 402 spectra, respectively, from the 600- and 400-line grism, and 
300-line grating.  The spectra were then separated into resolution groups and the 
algorithm was implemented independently on each sub-sample.  First, we searched 
between 1210 and 1225 \AA\ in each spectrum for the maximum flux value, given
that this wavelength range typically contains the observed Ly$\alpha$ peak in 
the spectra of star-forming galaxies.  This
maximum was identified as the primary Ly$\alpha$ emission peak.  The continuum level
was then estimated for the blue and red side separately from the mean of the flux 
over a $\sim$100 \AA\ span blueward and redward of 1210 and 1225 \AA, respectively.  
Starting from the primary peak's maximum flux point, we stepped down towards the 
blue side until reaching the continuum level.  If a local minimum was noted before 
reaching the continuum we marked the local maximum after that minimum as another 
possible peak for the Ly$\alpha$ line.  Another criterion used to avoid noise spikes 
was for the secondary maximum to span at least two adjacent steps in wavelength.  
This procedure was then repeated on the red side of the primary peak.  If an 
additional peak was noted by the algorithm, we required the flux ratio of the 
additional peak to the primary peak to be greater than $\frac{1}{N}$, with $N$ the 
S/N of the primary peak.  This value was chosen in order to reject any small noise 
spikes in the identification of the additional peak.  Our sample is therefore 
complete for primary to secondary ratios less than or equal to 3.  At higher 
primary-peak S/N ratio we can also probe larger ratios of primary-to-secondary peak 
heights.  In practice, however, the typical primary-to-secondary peak ratio is $\sim3$.
In addition to analyzing the full sample of spectra with our objective criteria, 
we examined the spectra by eye for any objects that might have been missed by 
the algorithm described above.  The additional multiple-peaked objects found by 
eye comprise a small percentage of the total sample, roughly $\sim15\%$.

For the 300-line grating, the percentage of spectra with multiple-peaked Ly$\alpha$ 
emission identified is 23$\%$ (92 spectra).  For the 600- and 400-line grisms the 
values are, respectively, 27$\%$ (44 spectra) and 20$\%$ (103 spectra).  In addition 
to the main spectroscopic sample, a set of 121 objects was followed up with 400-line 
grism spectra using significantly longer exposure times ($7-13$ hours as opposed to 
1.5 hours), with correspondingly higher S/N (Bogosavljevic 2010, Ph.D Thesis).  In
this sample, 97 objects show Ly$\alpha$ emission, and 33$\%$ (32 spectra) of the emitters 
are identified as containing multiple-peaked Ly$\alpha$ emission using the criteria 
described above.  Higher resolution (i.e., 600-line grism) and S/N 
(i.e., deeper 400-line grism) spectra are more likely to have identifiable 
multiple-peaked Ly$\alpha$ emission if it is present.  Based on the 600-line and 
deep 400-line samples we therefore assert that the prevalence of multiple-peaked 
profiles among objects with Ly$\alpha$ emission is $\sim$30$\%$ in a sample of 
UV-selected star-forming galaxies at $z\sim2-3$.  This percentage references a S/N 
lower limit of 3 on the strongest flux peak and a resolution limit of 225 km 
s$^{-1}$.

In order to access the rest-frame optical nebular emission lines (H$\alpha$ at $z\sim2$ 
and [OIII]$\lambda$5007 at $z\sim3$) enabling the determination of systemic velocities for
our galaxies, we are restricted to those objects in the full multiple-peaked sample 
whose redshifts ensure that either H$\alpha$ or [OIII]$\lambda$5007 falls into a window 
of atmospheric transmission.  The sharply increasing thermal background past 2.3 
$\mu$m further limits the redshift range over which rest-frame optical emission 
lines are accessible.  In the $\it{K}$-band the available redshift range is 
$z\sim2.0-2.6$ for H$\alpha$ and $z\sim2.9-3.7$ for [OIII].  In the $\it{H}$-band 
[OIII] can be accessed from $z\sim1.9-2.6$.  After applying the restricted redshift 
ranges to the multiple-peaked objects determined from the algorithm described above, 
we identified 193 objects as potential targets for near-infrared spectroscopic 
follow-up.  From this sample we selected the 18 objects
described in this paper, which are representative in terms of their Ly$\alpha$ kinematics (see Section \ref{sec:lyavel}).

\section{Observations and Data Reduction}
\label{sec:obsdata}

Both near-IR and optical spectra are required for our analysis
of the kinematics of Ly$\alpha$ emission profiles. In this section,
we describe both types of spectroscopic data, as well as our basic
empirical measurements and uncertainties.
 
\subsection{NIRSPEC Observations and Data Reduction}
\label{sec:nirspecobs}

Near-IR spectra were obtained using the NIRSPEC instrument \citep{mclean1998} on the 
Keck II telescope.  Our data were collected during three observing runs in 2009 
July, August, and October, for a total of 5 nights.  On average the total exposure 
time for each object was 4$\times$900 seconds, though the number of 900 second 
exposures per target ranged from 3 to 6 for a total integration time of 0.75 to 1.5 
hours.  All targets were observed with a $0.^{\prime\prime}76$ $\times$ 
$42^{\prime\prime}$ long slit.  For most objects we used the N6 filter, a broad 
$\it{H+K}$ filter centered at 1.925$\mu$m with a bandwidth of 0.75$\mu$m.  For 
objects at $z\sim2.5-2.6$, we used the N7 filter, which is centered at 2.23$\mu$m 
with a bandwidth of 0.80$\mu$m.  The spectral resolution as determined from sky 
lines was $\sim15$ \AA, which corresponds to $\Delta\upsilon\sim$230 km 
s$^{-1}$ (R$\simeq$1300) for the N6 filter and $\Delta\upsilon\sim$200 km 
s$^{-1}$ (R$\simeq$1500) for the N7 filter.  Conditions were photometric, with the 
seeing ranging from $0.^{\prime\prime}33 - 0.^{\prime\prime}70$.

Due to the faint nature of these objects in the $\it{K}$-band we acquired each 
target using blind offsets from a bright star in the surrounding field.  We returned 
to the offset star between each integration of the science target to recenter and 
dither along the slit.  In nearly all cases we dithered back and forth between two 
positions near the center of the slit.  The position angle was chosen to match that 
of the LRIS observation unless the object appeared extended at a different angle in 
the existing ${\cal R}$-band images \citep{steidel2003,steidel2004}.  We measured 
H$\alpha$ or [OIII]$\lambda$5007 for a total of 18 objects during our observing 
runs, which were drawn from the parent sample discussed in Section \ref{sec:samp}.  
All observational information can be found in Table \ref{tab:obs}.  $\cal R$-band 
images of the objects with the NIRSPEC and LRIS slits overlaid are presented in Figure 
\ref{fig:spec1}.

\begin{deluxetable*}{lccccccr}
\tablewidth{0pt} \tabletypesize{\footnotesize}
\tablecaption{Observation Log with Keck~II
NIRSPEC\label{tab:obs}}
\tablehead{
\colhead{Object} &
\colhead{R.A. (J2000)} &
\colhead{Dec. (J2000)} &
\colhead{$z_{\rm HII}$$^{a,b}$} &
\colhead{~$\cal{R}$(mag)~} &
\colhead{Exposure (s)} &
\colhead{Filter} &
\colhead{Date (UT)} 
}
\startdata
Q0100-MD40 & 1 03 18.971 &  13 17 02.841 & 2.2506 & 24.35 & 4 $\times$ 900 & N6 & Aug. 09, 2009 \\
Q0100-BX288 & 1 03 20.931 & 13 16 23.569 & 2.1032 & 23.75 & 2 $\times$ 900 & N6 & Aug. 09, 2009 \\
Q0449-BX167 & 4 52 24.624 & -16 39 15.284 & 2.3557	& 24.02 & 4 $\times$ 900 & N6 & Oct. 10, 2009 \\
Westphal-DD29 & 14 17 25.936 & 52 29 32.219 & 3.2401 & 24.82 & 4 $\times$ 900 & N6 & Aug. 10, 2009 \\
Westphal-BX154 & 14 17 32.173 & 52 25 51.044 & 2.5954 & 23.96 & 3 $\times$ 900 & N7 & July 02, 2009 \\
Q1549-BX102 & 15 51 55.977 & 19 12 44.220 & 2.1932 & 24.36 & 3 $\times$ 900 & N6 & July 01, 2009 \\
Q1549-C20 & 15 52 00.396 & 19 08 40.773 & 3.1174 & 24.87 & 3 $\times$ 900 & N6 & Aug. 10, 2009 \\
Q1549-M22$^c$ & 15 52 02.703 & 19 09 40.011 & 3.1535 & 24.85 & 1 $\times$ 900 & N6 & Jul. 03, 2009 \\
&&&&& 2 $\times$ 900 & N6 & Aug. 10, 2009\\
Q1549-BX202 & 15 52 05.090 & 19 12 49.818 & 2.4831 & 24.37 & 4 $\times$ 900 & N6 &  July 01, 2009\\
Q1623-BX129 & 16 25 28.732 & 26 49 19.182 & 2.3125 & 24.04 &  5 $\times$ 900 & N6 & July 02, 2009 \\
Q2206-BX151& 22 08 48.650 & -19 42 25.569 & 2.1974 & 24.03 & 4 $\times$ 900 & N6 & July 01, 2009 \\
SSA22a-D17$^d$ & 22 17 18.845 & 0 18 16.667 & 3.0851 & 24.27 & 6 $\times$ 900 & N6 & July 02, 2009 \\
SSA22a-C49$^d$ & 22 17 19.810 & 0 18 18.729 & 3.1531 & 23.85 & 6 $\times$ 900 & N6 & July 02, 2009 \\
SSA22a-C31 & 22 17 22.885 & 0 16 09.492 & 3.0176 & 24.61 & 3 $\times$ 900 & N6 & July 03, 2009 \\
SSA22a-C32 & 22 17 25.629 & 0 16 13.095 & 3.2926 & 23.68 & 4 $\times$ 900 & N6 & July 03, 2009 \\
Q2343-BX70 & 23 45 55.934 & 12 44 50.317 & 2.4086 & 24.94 & 4 $\times$ 900 & N6 & Aug. 09, 2009 \\
Q2343-BX231 & 23 46 20.108 & 12 46 16.812 & 2.4999 & 25.15 & 3 $\times$ 900 & N6 & Aug. 10, 2009 \\
Q2346-BX181 & 23 48 31.827 & 0 21 39.077 & 2.5429 & 23.28 &  3 $\times$ 900 & N7 & July 01, 2009 \\
\enddata
\tablecomments{Units of right ascension are hours, minutes, and seconds, and units of declination are degrees, arcminutes and arcseconds.}
\tablenotetext{a}{Rest-frame optical nebular emission-line redshifts.  Objects with $z\sim2.0-2.6$ have H$\alpha$ measurements. Objects with $z>3$ have [OIII]$\lambda$5007 measurements.}
\tablenotetext{b}{Typical uncertainty in $z_{\rm HII}$ is $\Delta z\sim10^{-4}$.}
\tablenotetext{c}{Object Q1549-M22 was observed on both July 3$^{\mbox{rd}}$ and August 10$^{\mbox{th}}$ and the science images from these nights were stacked together.}
\tablenotetext{d}{Objects SSA22a-D17 and SSA22a-C49 were obtained on the same slit.}
\end{deluxetable*}

Data reduction was performed following the method described in \citet{liu2008}, 
where the sky background was subtracted from the two-dimensional unrectified science 
images using an optimal method \citep[][G. D. Becker 2006, private
communication]{kelson2003}. After background subtraction, cosmic rays and bad pixels were 
removed from each exposure, which was then rotated, cut out along the slit, and 
rectified in two dimensions to take out the curvature both in the wavelength and 
spatial directions.  The sky lines were fit with a low-order polynomial and a 
b-spline fit was used in the dispersion direction.  The final rectified 
two-dimensional exposures for each object were then registered and combined into one 
spectrum.  A one-dimensional spectrum was extracted from the two-dimensional reduced 
image along with the corresponding 1$\sigma$ error spectrum.  The average aperture 
size along the slit was 2.$^{\prime\prime}$1, with a range of 1.$^{\prime\prime}$7 
to 2.$^{\prime\prime}$9.  The spectrum was then flux-calibrated using A-type stars 
observed during our NIRSPEC runs according to the method described in 
\citet{shapley2005} and \citet{erb2003}.  Finally the flux-calibrated, 
one-dimensional spectrum was placed in a vacuum, heliocentric frame.

\subsubsection{Objects With Spatially-Resolved NIRSPEC Spectra}
\label{sec:resolved}

\begin{figure}
\begin{center}
\centerline{
%    \mbox{\includegraphics[scale =0.75]{kinematics_xfig.eps}}
    \mbox{\includegraphics[scale =0.38]{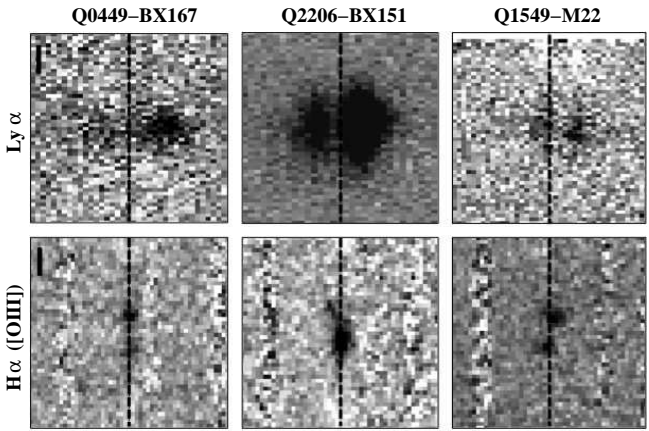}}
  }
\caption{From left to right, the two-dimensional LRIS and NIRSPEC spectra for 
Q0449-BX167, Q2206-BX151, and Q1549-M22.  These three objects exhibit 
spatially-extended nebular emission.  The horizontal and vertical axes represent, 
respectively, velocity with respect to the velocity zeropoint and spatial distance along the slit.
In each image, the dashed vertical line represents the systemic velocity of the system ($v=0$), and the horizontal axis spans $\pm 1,500$ km s$^{-1}$ in velocity. The vertical axes span $\pm 3.5$$^{\prime\prime}$ in the spatial direction, and a vertical scale bar of $1^{\prime\prime}$ is shown
in the upper left-hand panels for Q0449-BX167.  (Top) LRIS 2D Ly$\alpha$ spectra. (Bottom) NIRSPEC 2D spectra.  The spectra of Q0449-BX167 and Q2206-BX151 show the H$\alpha$ feature, while the spectrum of Q1549-M22 is centered on [OIII]$\lambda$5007.  Q0449-BX167 has two components separated by 
$1.^{\prime\prime}3$ along the slit, Q2206-BX151 is extended spatially by $1.^{\prime\prime}3$, and Q1549-M22 consists of two components separated by $1.^{\prime\prime}1$ in the spatial direction and 130 km s$^{-1}$ in the spectral direction.
\label{fig:2dspec}}
\end{center}
\end{figure}

While the majority of the objects in our sample have spatially unresolved H$\alpha$ 
(or [OIII]$\lambda$5007) emission-line spectra, three objects exhibit 
spatially-extended nebular emission: Q0449-BX167, Q2206-BX151, and Q1549-M22.  If 
spatially-extended emission is coupled with rotation or velocity shear, the 
placement or position angle of the long slit has the potential to yield nebular 
emission centroid and velocity dispersion values that differ from the true 
properties of the galaxy.  The differences may result if the slit does not evenly 
sample both the approaching and receding parts of the rotation curve or if 
systematic shear across the galaxy is interpreted as purely random motions.  The 
three objects listed above were analyzed in more detail to determine whether their 
complexity might have affected our measurements.  In the $\cal R$-band image for 
Q0449-BX167 there are two distinct components and our slit PA was chosen to have 
both components included.  Q0449-BX167 has two separate components in the 
two-dimensional NIRSPEC spectrum that are offset in the spatial direction by 
$1.^{\prime\prime}3$, but not in the spectral direction.  In the $\cal R$-band 
image, Q2206-BX151 is extended and our slit PA was chosen to encompass the extended 
region. In the two-dimensional NIRSPEC spectrum Q2206-BX151 is extended in the 
spatial direction for a total of $1.^{\prime\prime}3$.  There is a small tail that 
extends off of the main nebular emission, which is slightly offset in the spectral 
direction.  The flux from the tail is comparable to the noise seen in the image and 
was not extracted.  Since the structure in the two-dimensional spectra for 
Q0449-BX167 and Q2206-BX151 (excluding the tail) is not shifted in the spectral 
direction there is no significant evidence for rotation or shear, and the nebular 
emission centroid and dispersion values for these objects should be robust.  In 
fact, Q0449-BX167 and Q2206-BX151 have among the smallest velocity dispersion values 
of the sample.  The third object, Q1549-M22, is slightly extended in the $\cal 
R$-band image and again we chose our slit PA to correspond to the extended region.  
There are two distinct components in the two-dimensional spectrum for Q1549-M22 and 
they are offset by $1.^{\prime\prime}1$ and 9 \AA, respectively, 
in the spatial and spectral directions. 
The two-dimensional NIRSPEC spectra for these three objects are 
presented, along with their corresponding two-dimensional LRIS spectra, in Figure 
\ref{fig:2dspec}.

Object Q0449-BX167 was originally extracted to include only the brighter component 
on the left hand side of the two-dimensional spectrum.  To determine if the 
spatially-extended nebular emission affects our measurements, we re-extracted 
Q0449-BX167 to include both components seen in the two-dimensional spectrum.  For 
Q0449-BX167 the centroid did not change and the velocity dispersion changed by 
$\Delta\sigma_{v}$ = +2.5 km s$^{-1}$ when comparing the more extended extraction to 
our original one. This difference is within our errors.  For our analysis we utilize 
the values obtained from our original extraction of Q0449-BX167.  Object Q2206-BX151 
was originally extracted to include only the main elongated component.  We re-extracted 
to include the tail for comparison.  The centroid of the re-extraction did not 
change compared to the centroid measured from the original extraction.  The velocity 
dispersion changed from an upper limit of 41 km s$^{-1}$ (not including the tail) to 
a measurement of 63 km s$^{-1}$ (including the tail).  The re-extractions show that the centroid 
measurements used in our analysis for Q0449-BX167 and Q2206-BX151 are robust.  We 
include the velocity dispersion values measured from the original extraction for 
Q2206-BX151 in our analysis with the caveat that the added complexity might affect 
our modeling comparison (see Section \ref{sec:model}).

Our fiducial measurements for Q1549-M22 are based on the extraction of both spectral 
components.  Further analysis was conducted to understand how the spectral offset 
between the two components might affect our measurements.  Each component was 
extracted and reduced separately for comparison with the integrated extraction of 
both components.  Relative to the velocity zeropoint measured from the summed 
components, the offsets measured for each component separately are $\Delta v=110$ km 
s$^{-1}$ and $\Delta v=-76$ km s$^{-1}$.  The measured velocity dispersion of the 
spectrum integrated over both components (153 km s$^{-1}$) is approximately equal to 
the sum in quadrature of the velocity dispersion of the individual components (125 
km s$^{-1}$ and 95 km s$^{-1}$).  The shift from the velocity zeropoint of each 
component and the large measured velocity dispersion from the summed spectra (the 
largest in our sample) may be evidence of two separate components giving rise to the 
two Ly$\alpha$ peaks for this object.  We decided to include Q1549-M22 in our sample 
with the caveat that its complexity in both [OIII]$\lambda$5007 and Ly$\alpha$ may 
signify a different underlying phenomenon giving rise to the multiple-peaked 
Ly$\alpha$ emission from the one we will discuss in subsequent sections.

\subsection{LRIS Data Reduction}
\label{sec:lrisobs}

The observations and reduction of the LRIS spectra for our target sample have been 
described in previous work \citep{steidel2003,steidel2004}. 
These spectra were collected over the course of many years without consistent 
attention to the accuracy of the zeropoint of the wavelength solution at the 
50 to 100 km s$^{-1}$ level, given that their primary purpose was for 
the measurement of redshifts and basic galaxy properties.
In contrast, for the work presented here it is crucial to obtain a uniform and highly 
accurate wavelength zeropoint in order to place our LRIS and NIRSPEC data sets on 
the same velocity scale. Accordingly, we re-calculated the wavelength solution of each 
LRIS spectrum.  After computing the wavelength solution using the $\tt{IRAF}$ 
routine $\tt{identify}$ we adjusted the zeropoint of the solution such that bright, 
unblended sky lines (e.g., $\lambda$5577.339, $\lambda$6300.304) showed up at the 
correct wavelengths.  The spectra were then corrected from an observed, air frame to 
a heliocentric, vacuum one.

\subsection{Line Centroid, Flux, $\&$ FWHM Measurements}
\label{sec:lineflux}

Measuring an accurate nebular emission centroid in order to obtain a precise 
systemic redshift for our objects was the principal objective of the NIRSPEC 
observations described in Section \ref{sec:nirspecobs}.  The flux and FWHM were also 
important measurements for our analysis.  The lines that we set out to measure were 
H$\alpha$ at $z\sim2$ and [OIII]$\lambda$5007 at $z\sim3$.  The centroid, flux, and 
FWHM were determined by fitting a Gaussian profile to each emission line using the 
$\tt{IRAF}$ task, $\tt{splot}$.

A Monte Carlo approach was used to measure the uncertainties in the centroid, flux, 
and FWHM.  For each object 500 fake spectra were created by perturbing the flux at 
each wavelength of the true spectrum by a Gaussian random number with the standard 
deviation set by the level of the the 1$\sigma$ error spectrum.  Line measurements 
were obtained from the fake spectra in the same manner as the actual data. The 
standard deviation of the distribution of measurements from the artificial spectra 
was adopted as the error on each centroid, flux, and FWHM value.  The NIRSPEC 
measurements and associated uncertainties are given in Table \ref{tab:res1}.  The 
FWHM values listed are not corrected for instrumental broadening.  Based on H$\alpha$ or 
[OIII]$\lambda$5007 emission line centroid measurements, we obtained the systemic 
redshifts for our sample.  These redshifts were used to shift both the NIRSPEC and 
LRIS spectra into the systemic rest frame.  Based on the Monte Carlo uncertainty of 
the nebular emission line centroid or the RMS of the wavelength solution (depending 
on which was greater), we obtain an error of $\Delta\lambda\sim0.9$ \AA, which 
corresponds to a $\Delta v\sim10-15$ km s$^{-1}$.  Repeat observations of two 
objects from our sample using different slit PAs suggest a larger systematic 
uncertainty on the order of 50 km s$^{-1}$.  However, this value should be 
considered an upper limit on the redshift uncertainty given that we attempted to 
match the slit positions between NIRSPEC and LRIS observations for a significant 
fraction of our sample.

\begin{deluxetable}{lccr}
\tablewidth{0pt} \tabletypesize{\footnotesize}
\tablecaption{NIRSPEC Measurement Results\label{tab:res1}}
\tablehead{
\colhead{Object} &
\colhead{$z_{\rm HII}$$^a$} &
\colhead{FWHM$_{\rm HII}$$^{a,b}$} &
\colhead{$F_{\rm HII}$$^{a,c}$} 
}
\startdata
Q0100-MD40$^{d}$ & 2.2506 & 14.3 $\pm$ 1.0 & 3.6 $\pm$ 0.3 \\
Q0100-BX288 & 2.1032 & 19.6 $\pm$ 0.4 & 6.7 $\pm$ 0.1 \\
Q0449-BX167 & 2.3557 & 17.3 $\pm$ 1.4 & 7.8 $\pm$ 0.7 \\
Westphal-DD29 & 3.2401 & 22.9 $\pm$ 1.3 & 22.9 $\pm$ 0.5 \\
Westphal-BX154 & 2.5954 &  17.7 $\pm$ 2.0 & 7.7 $\pm$ 0.8 \\
Q1549-BX102 & 2.1932 & 18.2 $\pm$ 1.1 & 6.1 $\pm$ 0.3 \\
Q1549-C20 & 3.1174 & 17.3 $\pm$ 1.0 & 7.3 $\pm$ 0.4 \\
Q1549-M22 & 3.1535 & 29.2 $\pm$ 2.8 & 8.1 $\pm$ 0.6 \\
Q1549-BX202 & 2.4831 & 26.2 $\pm$ 9.9 & 4.2 $\pm$ 0.9 \\
Q1623-BX129 & 2.3125 & 21.1 $\pm$ 2.9 & 3.5 $\pm$ 0.4 \\
Q2206-BX151$^{e}$ & 2.1974 & 15.7 $\pm$ 0.7 & 6.3 $\pm$ 0.3 \\
SSA22a-D17 & 3.0851 & 25.1 $\pm$ 2.5 & 2.6 $\pm$ 0.3 \\
SSA22a-C49 & 3.1531 & 18.0 $\pm$ 0.6 & 6.4 $\pm$ 0.2 \\
SSA22a-C31 & 3.0176 & 19.7 $\pm$ 1.5  & 11.5 $\pm$ 0.9 \\
SSA22a-C32$^{d}$ & 3.2926 & 12.8 $\pm$ 35 & 5.2 $\pm$ 4.7 \\
Q2343-BX70$^{e}$ & 2.4086 & 16.6 $\pm$ 3.5 & 4.0 $\pm$ 0.6 \\
Q2343-BX231& 2.4999 & 21.5 $\pm$ 1.1 & 7.2 $\pm$ 0.4 \\
Q2346-BX181 & 2.5429 & 28.3 $\pm$ 2.6 &13.8 $\pm$ 1.2 \\
\enddata
\tablenotetext{a}{Rest-frame optical nebular emission-line redshifts.  Objects with 
$z\sim2.0-2.6$ have H$\alpha$ measurements. Objects with $z>3$ have 
[OIII]$\lambda$5007 measurements.}
\tablenotetext{b}{Observed FWHM and error in units of \AA.  The FWHM is a raw value, 
uncorrected for instrumental broadening.}
\tablenotetext{c}{Emission-line flux and error in units of $10^{-17}$ ergs s$^{-1}$ 
cm$^{-2}$.}
\tablenotetext{d}{Measured FWHM is less than instrumental resolution.}
\tablenotetext{e}{Measured FWHM is not appreciably higher than the instrumental 
resolution.}
\end{deluxetable}

\section{Ly$\alpha$ Velocity Profiles}
\label{sec:lyavel}

\begin{figure*}[htbp]
\begin{center}
\centerline{
%    \mbox{\includegraphics[scale =1.0]{vel_all_zero.eps}}
    \mbox{\includegraphics[scale = 0.9]{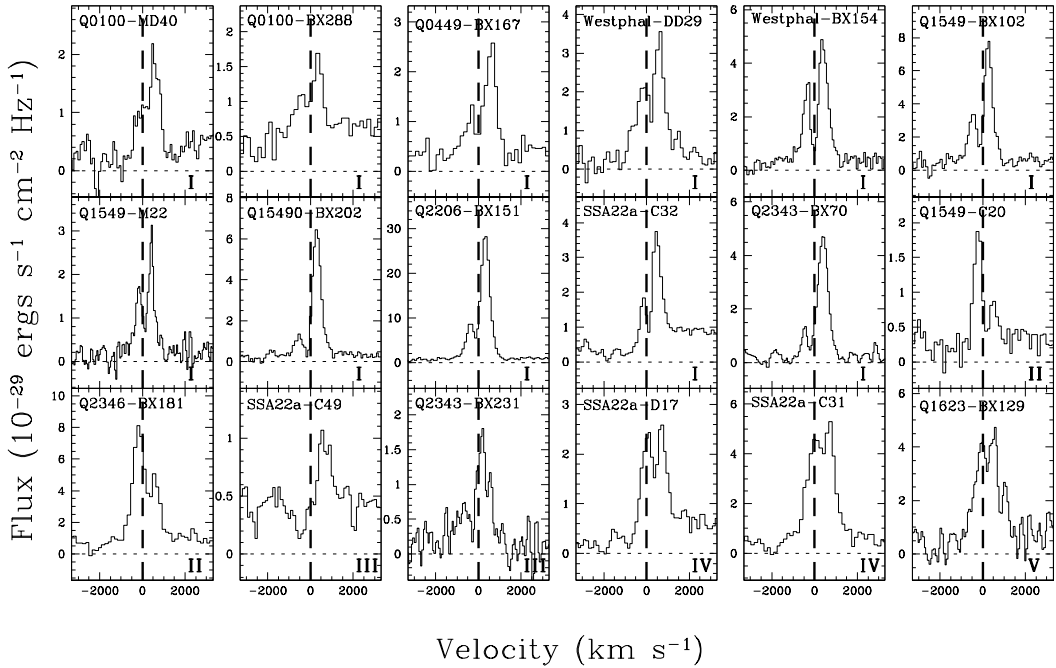}}
  }
\caption{Ly$\alpha$ velocity profiles for the 18 objects in our sample. The 
zeropoint velocity as measured from the nebular emission (H$\alpha$ at $z\sim2$ or 
[OIII]$\lambda$5007 at $z\sim3$) is indicated on each image with a dashed line.  
Considerable variety is evident among the line profile peaks and their ratios.  From 
top to bottom and left to right, the panels proceed from Group I to Group V 
profiles.
\label{fig:vel}}
\end{center}
\end{figure*}

Multiple-peaked Ly$\alpha$ emission profiles in $z\sim 2-3$ UV-selected galaxies
in principle encode a wealth of information about the velocity and density fields 
through which they have propagated.  It is known that many galaxies at this redshift range are 
experiencing outflows \citep{shapley03,veilleux05}.  Simple galactic 
wind models that incorporate radiative transfer suggest that the specific velocity
profile of the resultant Ly$\alpha$ line $\it{relative~to~systemic}$ can be used to 
determine parameters of the wind such as the expansion velocity, column density of 
windswept material, and outflow temperatures \citep{verhamme06,verhamme08}.  With 
our measured systemic redshifts and Ly$\alpha$ morphologies, we are now able to 
analyze our systems with this comparison in mind.  After being shifted
to the systemic rest frame based on H$\alpha$ or [OIII] redshifts, the LRIS Ly$\alpha$ 
spectra were translated from wavelength to velocity space using the rest wavelength 
of the Ly$\alpha$ feature to create the Ly$\alpha$ velocity profiles.  The velocity 
profiles are shown in Figure \ref{fig:vel} and the corresponding velocity measurements for each peak and trough are shown in Table \ref{tab:vel}.

\begin{figure}
\begin{center}
\centerline{
 %   \mbox{\includegraphics[scale =0.80]{Group_xfig.eps}}
      \mbox{\includegraphics[scale =0.43]{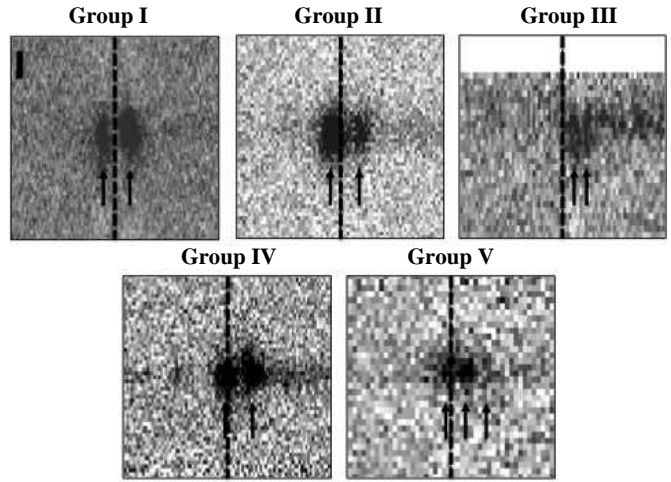}}
  }
\caption{ Two-dimensional Ly$\alpha$ emission spectra representing each of our 5 
groups.  For each image the horizontal axis represents velocity with respect to the velocity zeropoint, where the dashed vertical line represents the systemic velocity ($v=0$).  The horizontal axis in each panel spans $\pm3,000$ km s$^{-1}$.  The vertical axis represents spatial distance along the slit, with a span of
$\pm 3.5^{\prime\prime}$, and a vertical scale bar of $1^{\prime\prime}$ is shown in the upper left-hand panel (labeled Group I).  Group I (two peaks that straddle the velocity-field zeropoint, with the stronger flux peak on the red side) is represented by Westphal-BX154, Group II (two peaks that straddle the velocity-field zeropoint, with the stronger flux peak on the blue side) is represented by Q2346-BX181, Group III (both peaks are shifted completely redward of the zeropoint, with the stronger flux peak on the blue side) 
by SSA22a-C49, Group IV (two peaks that straddle the velocity-field zeropoint with 
both peaks having approximately the same flux with the red peak marginally stronger) 
by SSA22a-D17, and Group V (three peaks) by Q1623-BX129. In each panel, the locations of the Ly$\alpha$ peaks are indicated by arrows.
\label{fig:2Dlya}}
\end{center}
\end{figure}

\begin{deluxetable*}{lccccccr}
\tablewidth{0pt} \tabletypesize{\footnotesize}
\tablecaption{Velocity Measurements from Ly$\alpha$ Emission and Interstellar Absorption\label{tab:vel}}
\tablehead{
\colhead{Object} &
\colhead{$z_{\rm HII}$$^{a}$} &
\colhead{First Peak$^b$} &
\colhead{Second Peak$^b$} &
\colhead{Trough$^b$} &
\colhead{Absorption$^b$} &
\colhead{Grism/Grating$^c$}
}
\startdata
Q0100-MD40$^{d}$ & 2.2506 & $-461$ & $+550$ &$+288$ & $-19$&600, 400\\
Q0100-BX288 & 2.1032 & $-579$ & $+377$ & $-104 $ & $-317$&400\\
Q0449-BX167 & 2.3557 & $-348$ &$+586$ & $-60$ & $-47$&400\\
Westphal-DD29 & 3.2401 & $-259$ &$+471$ & $+202$ & $\cdot \cdot \cdot$&300\\
Westphal-BX154 & 2.5954 & $-353 $ &$+404$ &  $-27$ & $-204$&600\\
Q1549-BX102 & 2.1932 &  $-478$ & $+264$ & $-155$ & $\cdot \cdot \cdot$&600 \\
Q1549-C20 & 3.1174 & $-237$ & $+533$ &  $+210$ & $+14$&400\\
Q1549-M22 & 3.1535 & $-153$ & $+414$ & $+111$ & $\cdot \cdot \cdot$&600\\
Q1549-BX202 & 2.4831 & $-542$ &  $+301$ & $-141$ & $-143$&600\\
Q1623-BX129$^{e}$ & 2.3125 & $-42$ &  $+542$ & $+229$ & $-468$&400\\
Q2206-BX151 & 2.1974 & $-417$ &  $+328$ & $-148$ & $\cdot \cdot \cdot$&600\\
SSA22a-D17 & 3.0851 & $-25$ & $+732$ & $+385$ & $-1174$&400\\
SSA22a-C49 & 3.1531 & $+518$ &  $+903$ & $+735$ & $-232$&400\\
SSA22a-C31 & 3.0176 & $-15$ & $+695$ &  $+424$ & $\cdot \cdot \cdot$&300\\
SSA22a-C32 & 3.2926  & $-155$ & $+493$ &  $+99$ & $-24$&400\\
Q2343-BX70 & 2.4086 & $-476$ & $+399$ &  $-214$ & $-33$&400\\
Q2343-BX231& 2.4999 & $+84$ &  $+567$ & $+399$ & $\cdot \cdot \cdot$&400\\
Q2346-BX181 & 2.5429 & $-214$ & $+626$ & $+338$ & $-49$&400\\
\enddata
\tablenotetext{a}{Rest-frame optical nebular emission-line redshifts.  Objects with 
$z\sim2.0-2.6$ have H$\alpha$ measurements. Objects with $z>3$ have 
[OIII]$\lambda$5007 measurements.}
\tablenotetext{b}{Velocity relative to systemic in units of km s$^{-1}$.}
\tablenotetext{c}{All Ly$\alpha$ spectra were obtained on the LRIS spectrograph at 
Keck Observatory.  These spectra were taken with the 300 line mm$^{-1}$ grating 
blazed at 5000 \AA, or the 400 line mm$^{-1}$ or 600 line mm$^{-1}$ grism blazed at 
3400 \AA, and 4000 \AA, respectively, following the LRIS-B upgrade 
\citep{steidel2004}.}
\tablenotetext{d}{Object Q0100-MD40 had two spectra taken at separate observing 
runs, with separate gratings.  We combined both to make the final spectrum.}
\tablenotetext{e}{Object Q1623-BX129 contains a third peak.  The corresponding 
velocity is $+1051$ km s$^{-1}$, while the second trough is at $+846$ km s$^{-1}$.}
\end{deluxetable*}

The Ly$\alpha$ velocity profiles can be separated into groups characterized by the 
strengths and locations of the emission peaks with respect to the velocity-field 
zeropoint.  The majority of objects in our sample (11/18), which we refer to as 
Group I, have two peaks that straddle the velocity-field zeropoint, with the 
stronger flux peak on the red side.  This group includes objects Q0100-MD40, 
Q0100-BX288, Q0449-BX167, Westphal-DD29, Westphal-BX154, Q1549-BX102, Q1549-M22, 
Q1549-BX202, Q2206-BX151, SSA22a-C32, and Q2343-BX70.  In the case of Q1549-M22, the object 
with two apparently distinct spectral and spatial [OIII] emission components, 
regardless of whether we shift the velocity-field zeropoint to $\Delta v=110$ km 
s$^{-1}$ or $\Delta v=-76$ km s$^{-1}$, as measured from the re-extraction of each 
component individually, both Ly$\alpha$ peaks still straddle the zeropoint.  Therefore 
Q1549-M22 is classified as a Group I object independent of the adopted velocity zeropoint.
Group II spectra, including objects Q1549-C20 and Q2346-BX181, are described by two 
peaks that straddle the velocity-field zeropoint, and have the stronger flux peak on 
the blue side.  Objects Q2343-BX231 and SSA22a-C49 comprise Group III, where the 
bluer Ly$\alpha$ peak is stronger, but unlike the profiles previously described, 
both peaks are shifted completely redward of the zeropoint.  Group IV, SSA22a-D17 
and SSA22a-C31, have two peaks that straddle the velocity-field zeropoint with both 
peaks having approximately the same flux with the red peak marginally stronger.  
Object Q1623-BX129 accounts for the last group, Group V.  This object has three 
peaks.  Two of the peaks are similar to those of SSA22a-D17 and SSA22a-C31, with 
nearly the same flux and one on either side of the zeropoint. The third peak, which 
is furthest redward from the zeropoint, has the smallest flux.  Examples of 
two-dimensional Ly$\alpha$ spectra from each group are displayed in Figure 
\ref{fig:2Dlya}.

\begin{figure}
\begin{center}
\centerline{
%    \mbox{\includegraphics[scale =0.80]{velocity_peaks_all.eps}}
    \mbox{\includegraphics[scale =0.44]{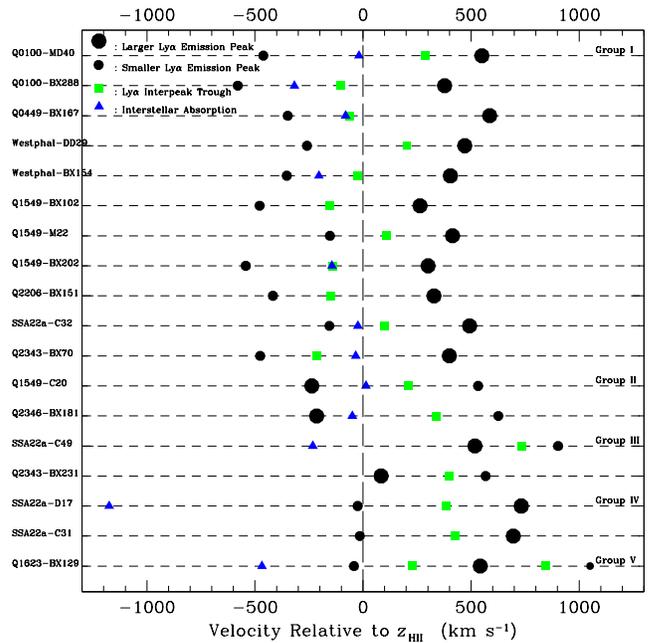}}
  }
\caption{The location of the velocity for every peak, trough, and absorption 
line with respect to the velocity-field zeropoint.  From 
top to bottom, the properties are displayed for Group I, II, III, IV, and V objects.  
Objects are offset vertically for clarity.  In each line the larger [smaller] black dot 
represents the velocity peak with the larger [smaller] flux.  The green square represents the 
velocity of the trough between the peaks.  The blue triangle represents the 
absorption line velocity if it was measured for the object.  For Q1623-BX129, 
which has three measured peaks, an even smaller black dot is used to indicate
the third peak with the weakest flux.
\label{fig:vel2}}
\end{center}
\end{figure}

A precise velocity-field zeropoint in addition to the velocity dispersion (see 
Section \ref{sec:veldisp}) provides constraints on radiative transfer models of the 
Ly$\alpha$ emission from our sample of galaxies (see Section \ref{sec:model}).  
Figure \ref{fig:vel2} shows a graphical summary of peak velocities along with the 
relative fluxes of each peak for every galaxy in our sample.  Additionally, some 
objects have detectable interstellar absorption lines, which are also labeled on 
Figure \ref{fig:vel2}.  The analysis of the absorption lines is described in Section 
\ref{sec:abs}.  The velocity separation of the peaks in the Ly$\alpha$ profiles is 
also an important constraint on possible models.  The mean velocity peak 
separation is $\langle \Delta v_{peak} \rangle = 741 \pm 39$ km s$^{-1}$ 
(where the error represents the standard deviation of the mean),
and the median is $\Delta v_{peak,med}$ = 757 km s$^{-1}$.

\begin{figure*}[htbp]
\begin{center}
\centerline{
%    \mbox{\includegraphics[scale =0.5]{peak_peak.eps}}
%    \mbox{\includegraphics[scale =0.5]{peak_peak_parent.eps}}
    \mbox{\includegraphics[scale =0.4]{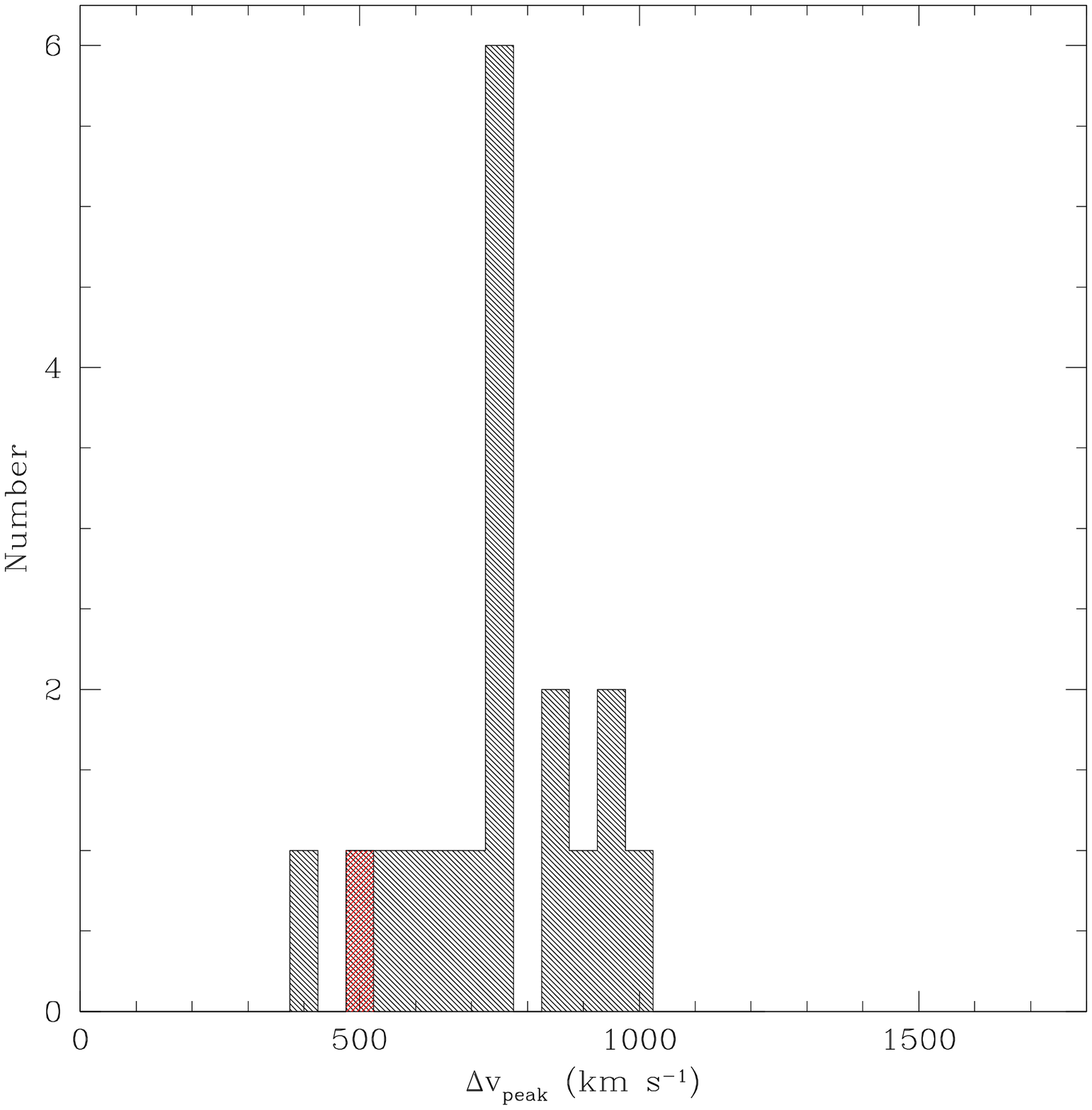}}
    \mbox{\includegraphics[scale =0.4]{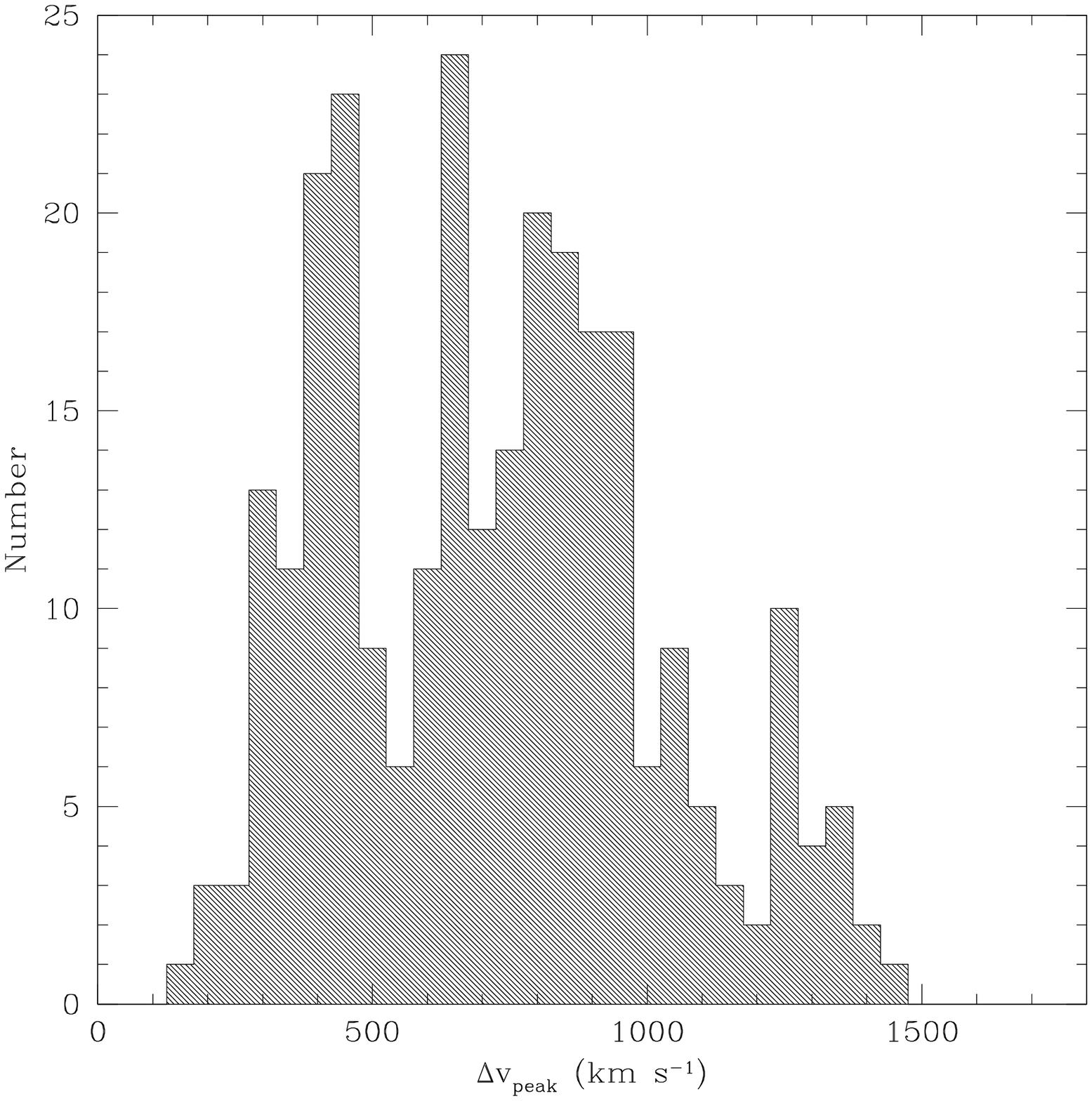}}
  }
\caption{Histogram of the distribution of peak-to-peak velocity separation.  (Left) 
The histogram for the 18 objects with NIRSPEC measurements.  The red shaded bar 
represents the separation of the 2nd peak (the center peak) from the 3rd peak (the 
reddest peak) for object Q1623-BX129.  The mean value for the sample is $\langle \Delta 
v_{peak} \rangle =741 \pm 39$ km s$^{-1}$, with a median of
$\Delta v_{peak,med}=757$~km s$^{-1}$. (Right) Histogram of the peak-to-peak 
separation for our parent LRIS sample with a S/N lower limit of 3.  The mean
value is $\langle \Delta v_{peak} \rangle = 723 \pm 18$~km s$^{-1}$, with a median
of $\Delta v_{peak,med}=721$ km s$^{-1}$.
\label{fig:separ}}
\end{center}
\end{figure*}

An examination of the parent LRIS multiple-peaked sample indicates the same trends 
as described above.  Analyzing the objects in the parent sample with a S/N lower 
limit of 3 on the strongest flux peak reveals that $\sim$65$\%$ have their strongest 
flux peak located on the red side of the interpeak trough (compared to 72$\%$ from 
the NIRSPEC sample).  Despite a wide range from minimum to maximum for the 
peak-to-peak velocity separation, the statistics for the parent LRIS sample are
comparable to those for our NIRSPEC sample, with a mean value of 
$\langle \Delta v_{peak} \rangle =$ 723 $\pm$ 18 km s$^{-1}$ 
(where the error represents the standard deviation of the mean), and
a median of $\Delta v_{peak,med} =$ 721 km s$^{-1}$.  
Figure \ref{fig:separ} shows the distributions of peak-to-peak velocity separation 
for our NIRSPEC sample and the parent LRIS sample.  These distributions demonstrate 
that our sample of 18 objects with NIRPSEC observations is representative of the 
parent LRIS sample as a whole.

\section{Physical Quantities}
\label{sec:physquant}

In this section, we discuss several relevant physical properties for the NIRSPEC multiple-peaked sample objects that can be estimated from our combined dataset.

\subsection{Velocity Dispersion}
\label{sec:veldisp}

As traced by H$\alpha$ (or [OIII]$\lambda$5007) emission, the nebular velocity 
dispersion, $\sigma_{v}$, reflects the integrated emission from star forming 
regions, which should reveal the potential well within the central few kiloparsecs 
of a high-redshift star-forming galaxy.  Since the H$\alpha$ (or 
[OIII]$\lambda$5007) velocity dispersion also traces the initial velocity dispersion 
of the Ly$\alpha$ photons, it is an important input parameter for models of 
Ly$\alpha$ radiative transfer and will enable a better understanding of the 
multiple-peaked Ly$\alpha$ emission systems in our sample.  In this paper, aside from a brief comparison with previous work (Section \ref{sec:modelcomp}), we do not include a detailed analysis of the relationship between intrinsic velocity dispersion and the emergent Ly$\alpha$ profile.  Such analysis is deferred to future work, as discussed in Section \ref{sec:conclusions}.

\begin{figure}
\begin{center}
\centerline{
%    \mbox{\includegraphics[scale =0.80]{sigma.eps}}
    \mbox{\includegraphics[scale =0.45]{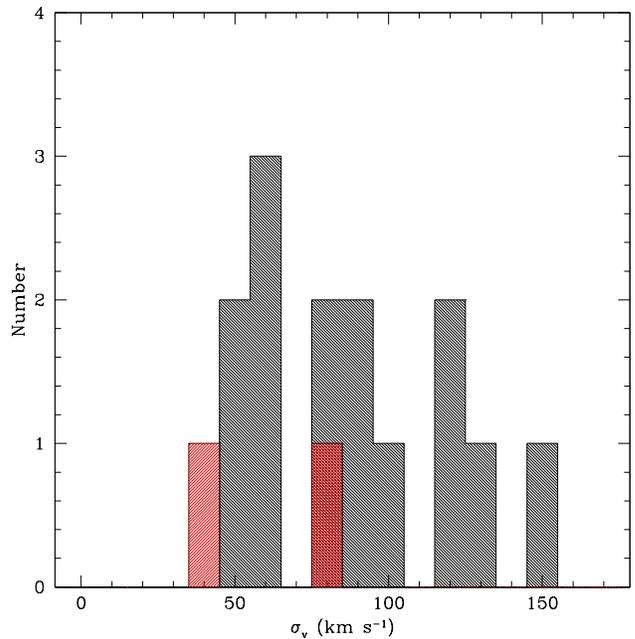}}
  }
\caption{Histogram of the nebular velocity dispersion, $\sigma_{v}$, as traced by 
H$\alpha$ or [OIII]$\lambda$5007 measurements.  The measurements for two objects with FWHM not 
appreciably larger than the instrumental resolution, Q2206-BX151 and Q2343-BX70, are 
shaded red.  The two objects that are considered unresolved, Q0100-MD40 and 
SSA22a-C32, are not included in this figure.  The mean value for the sample of
14 objects with measured velocity dispersions is $\langle \sigma_{v} \rangle =$ 90$\pm$ 9 km s$^{-1}$.
The median value for the full sample of 18 objects is $\sigma_{v,med}=$ 82 km s$^{-1}$.
\label{fig:sig}}
\end{center}
\end{figure}

The velocity dispersion was calculated as $\sigma_{v}$ $=$ (FWHM/2.355) $\times(c/\lambda_{obs})$, where 
$\lambda_{obs}$ is the observed wavelength of H$\alpha$ or [OIII]$\lambda$5007.  The 
FWHM is determined from the subtraction of the instrumental FWHM (estimated from the 
widths of night sky lines) from the observed FWHM in quadrature. Uncertainties in 
the velocity dispersion were determined using the same method described in 
\citet{erb2006b}.  Of our 18 objects, 14 have measurable velocity dispersion.  For 2 
additional objects we obtained an upper limit in the velocity dispersion because the 
measured line width was not appreciably higher than the instrumental resolution.  
Finally, 2 objects have measured line widths that are smaller than the instrumental 
resolution and are therefore considered unresolved.  Figure \ref{fig:sig} shows the 
distribution of the velocity dispersion for the sample.  The mean value for the
sample of 14 objects with measurable velocity dispersion is 
$\langle \sigma_{v}\rangle $ = 90$\pm$ 9 km s$^{-1}$ (where the error
represents the standard deviation of the mean), while the median for the
full sample of 18 objects is $\sigma_{v,med} $= 82 km s$^{-1}$. 
These are similar to the typical values 
reported by \citet{erb2006b} and \citet{pettini01} for $z\sim2-3$ UV-selected 
galaxies ($\sigma_{v}=108\pm5$ km s$^{-1}$ for $z\sim2$ and $\sigma_{v}=84\pm5$ km 
s$^{-1}$ for $z\sim3$).

\subsection{Dust Extinction\label{sec:dust}}

Ly$\alpha$ emission can be significantly suppressed due to attenuation by dust.  
Commonly, dust extinction in $z\sim2$ star-forming galaxies is estimated from 
rest-frame UV colors and an assumption of the \citet{calzetti2000} starburst 
attenuation law.  Alternatively, the amount of dust extinction can be estimated from 
the observed ratio of hydrogen Lyman to Balmer lines, since their intrinsic ratios 
are well described by atomic theory.  An assumption of Case B recombination and 
$T=10,000K$ determines the intrinsic Ly$\alpha$ to H$\alpha$ ratio 
\citep{osterbrock89}.  Any deviation from this value is then attributed to dust 
extinction.  In our sample, only 11 of 18 galaxies have measured H$\alpha$ (the 
remainder have [OIII]$\lambda$5007), making the first method more generally 
applicable for our analysis.  Both methods were used to calculate $E(B-V)$ for 
objects with measured H$\alpha$.

For the first method, we used the $G -\cal R$ colors of our sample as determined 
from \citet{steidel2003,steidel2004}.  If the Ly$\alpha$ line fell in the $G$-band 
(at $z\sim2.48-3.28$) the color was corrected for emission-line contamination using 
$\Delta m = 2.5 \times \log(1+\frac{EW}{\Delta G})$, where $\frac{EW}{\Delta G}$ 
is the ratio of the observed 
equivalent width of the Ly$\alpha$ emission feature to the bandwidth of the $G$ 
filter (1100 \AA).  The correction, $\Delta m$, was then added to the $G -\cal R$ 
color.  The $G -\cal R$ color for objects at $z>2.45$ was also corrected for IGM 
absorption in the $G$-band \citep{madau95}.  An intrinsic SED model was assumed as 
well to estimate the dust extinction.  We used a \citet{bruzchar03} model assuming 
solar metallicity and constant star formation rate with a stellar age of 570 Myr for 
$z\sim2$ and 202 Myr for $z\sim3$ objects.  Our choice of stellar age for objects at 
$z\sim2$ was based on results from \citet{erb2006b}, who found a median age of 570 
Myr for star-forming galaxies in this redshift range.  The stellar age of 202 Myr 
adopted for the $z\sim$3 galaxies was based on results from \citet{shapley2005}. The 
calculated values of $E(B-V)$ are listed in Table \ref{tab:physquant}.

For the second method, based on the ratio of Ly$\alpha$ and H$\alpha$ emission 
lines, we adopt the flux ratio $\frac{Ly\alpha_{(int)}}{H\alpha_{(int)}} = 8.7$ 
\citep{osterbrock89}.  We used the following equation:

\begin{equation}
\frac{Ly\alpha_{(obs)}}{H\alpha_{(obs)}} = 8.7\times10^{-0.4 E(B-V)(\kappa_{Ly\alpha} - \kappa_{H\alpha})}
\end{equation}

In this equation $\kappa_{Ly\alpha}$ and $\kappa_{H\alpha}$ refers to the selective 
extinction at the wavelengths of Ly$\alpha$ and H$\alpha$, respectively, as defined 
in Equation 4 of \citet{calzetti2000}.  $E(B-V)$ can then be calculated based on the 
observed ratio of Ly$\alpha$ to H$\alpha$ flux.  All of our calculated $E(B-V)$ 
values from this method exceed those measured from the rest-frame UV colors, and the 
median excess is a factor of 1.5.  This difference may be explained with reference 
to \citet{calzetti2000}, which states that the stellar continuum in local 
star-forming galaxies should suffer less reddening than the ionized gas.  
Furthermore, Ly$\alpha$ may suffer additional extinction due to resonant scattering 
and an effectively longer path length though the interstellar medium.  In practice, 
it is difficult to obtain a robust estimate of the Ly$\alpha$ to H$\alpha$ ratio due 
to the effects of differential slit loss.  \citet{erb2006c} find that NIRSPEC 
H$\alpha$ spectra using the 0.76$^{\prime\prime}$ slit may result in a typical 
factor of $\sim$2 loss, while \citet{steidel11} find that LRIS 1.2$^{\prime\prime}$ 
slit spectra of Ly$\alpha$ may result in typical losses of a factor of $\sim$5.  In 
detail, the relative loss of Ly$\alpha$ to H$\alpha$ depends on the spatial 
distribution of the emission in each of these lines.  Overall, the Ly$\alpha$ losses 
appear to outweigh those of the H$\alpha$ losses, which will tend to cause an 
underestimate in the Ly$\alpha$ to H$\alpha$ ratio.  Because of the systemic 
uncertainty of the Ly$\alpha$ to H$\alpha$ ratio and due to the fact that we can 
only estimate $E(B-V)$ for half of the sample using the (uncertain) intrinsic ratio 
method, we adopt the values calculated from the rest-frame UV colors.  The average 
value for our sample when employing the first method is $\langle E(B-V) \rangle $ = 0.09, which 
is bluer than the typical value seen at these redshifts, $\langle 
E(B-V)\rangle$ = 0.15 \citep{reddy08}.  This difference is not surprising 
given that our sample is entirely composed of objects with Ly$\alpha$ emission and 
given the apparent anti-correlation between Ly$\alpha$ emission EW and dust 
extinction \citep{shapley03,kornei10}.

\subsection{Star-Formation Rates and Intrinsic Luminosity}
\label{sec:sfr}

Star-formation rates were calculated from the rest-frame UV luminosity.  For sources 
at $z\sim2$ we used the $G$-band as a proxy for rest-frame UV luminosity.  At 
$z\sim3$, $\cal R$-band is used as an equivalent probe.  The observed magnitudes 
were first corrected for dust extinction using $m_{corr} = m_{app} - (\kappa \times 
E(B-V))$, where $\kappa$ is defined in Section \ref{sec:dust}.  Given the range of 
redshifts in our $z\sim2$ sample the $G$-band effective wavelength of 4780 \AA\ 
corresponds to a rest-frame wavelength range of $1330-154$0 \AA.  For the $z\sim3$ 
sample the $\cal R$-band effective wavelength translates to a rest-frame effective 
wavelength of $1590 - 1700$ \AA.  These rest-frame wavelengths were used to 
determine $\kappa$ values.  The corrected magnitude was then used to determine the 
flux, $F_{\nu}$.  Using the luminosity distance, we calculated the rest-frame UV 
luminosity density according to $L_{\nu}=(4 \pi {D_{L}}^{2} F_{\nu})/(1+z)$.  From 
\citet{kennicuttAR98}, the relation of $\mbox{SFR}(M_{\odot}$ yr$^{-1}) = 7.8 \times 
10^{-29} $ $L_{\nu}$(ergs s$^{-1}$ Hz$^{-1}$) was used to determine the SFRs 
assuming a \citet{chabrier03} IMF.  The SFRs for the sample range from 6$-$110 
$M_{\odot}$ yr$^{-1}$. The average value is 32 $M_{\odot}$ yr$^{-1}$, which is 
fairly typical of UV-selected star-forming galaxies at $z\sim2-3$ \citep{erb2006c}.  
The intrinsic Ly$\alpha$ luminosity, $L_{Ly\alpha}$, can be derived from the SFR 
using the conversion described in \citet{kennicutt98} and again assuming a \citet{chabrier03} IMF:

\begin{equation}
L_{Ly\alpha}(\mbox{ergs}~ \mbox{s}^{-1}) = \frac{\mbox{SFR}~(M_{\odot} ~\mbox{yr}^{-1})}{5.1 \times 10^{-43}}
\end{equation}

Based on the average SFR of 32 $M_{\odot}$ yr$^{-1}$, the intrinsic Ly$\alpha$ 
luminosity is $\langle L_{Ly\alpha} \rangle= 6.3 \times 10^{43}$ ergs s$^{-1}$.  The intrinsic 
Ly$\alpha$ luminosity is an important quantity for our models because it allows us 
to input an accurate description of the photon source from our galaxies.  SFRs and 
$L_{Ly\alpha}$ values are listed in Table \ref{tab:physquant}.

\begin{deluxetable}{lcccr}
\tablewidth{0pt} \tabletypesize{\footnotesize}
\tablecaption{Physical Quantities\label{tab:physquant}}
\tablehead{
\colhead{Object} &
\colhead{$\sigma_{v}~^{a}$} &
\colhead{$E(B-V)$}$^{b}$ &
\colhead{SFR$^{c}$} &
\colhead{$L_{{\rm Ly}\alpha}$$^{d}$}
}
\startdata
Q0100-MD40 & $\cdot \cdot \cdot$ & 0.00 & 6 & 1.25\\
Q0100-BX288 & 84.8$^{+3}_{-3}$ & 0.02 & 11 & 2.16\\
Q0449-BX167 & 50.3$^{+14}_{-19}$ & 0.20 & 51 & 10.10\\
Westphal-DD29 & 102.5$^{+10}_{-11}$ & 0.00 & 73 & 14.90\\
Westphal-BX154 & 52.4$^{+18}_{-24}$ & 0.26 & 110 & 21.70\\
Q1549-BX102 & 64.6$^{+10}_{-13}$ & 0.10 & 13 & 2.51\\ 
Q1549-C20 & 55.2$^{+11}_{-13}$ & 0.04 & 9 & 1.83\\
Q1549-M22 & 153.1$^{+19}_{-20}$ & 0.15 & 27 & 5.23\\
Q1549-BX202 & 119.7$^{+63}_{-84}$ & 0.07 & 14 & 2.78\\
Q1623-BX129 & 89.5$^{+22}_{-25}$ & 0.00 & 9 & 1.74\\
Q2206-BX151 & $<$ 40.5 & 0.01 & 8 & 1.66 \\
SSA22a-D17 & 124.9$^{+19}_{-20}$ & 0.08 & 24 & 4.84\\
SSA22a-C49 & 60.5$^{+6}_{-7}$ & 0.07 & 33 & 6.61\\
SSA22a-C31 & 82.0$^{+14}_{-15}$ & 0.00 & 8 & 1.59\\
SSA22a-C32 & $\cdot \cdot \cdot$ & 0.04 & 31 & 6.19\\
Q2343-BX70 & $<$ 78.5 & 0.13 & 12 & 2.35\\
Q2343-BX231& 86.3$^{+8}_{-9}$ & 0.29 & 47 & 9.20\\
Q2346-BX181 & 131.8$^{+17}_{-17}$ &  0.18 & 97 & 19.20\\
\enddata
\tablenotetext{a}{Velocity dispersion measured in units of km s$^{-1}$.}
\tablenotetext{b}{$E(B-V)$ value derived from $G-\cal{R}$ color.}
\tablenotetext{c}{Star-formation rate measured in units of M$_{\odot}$ yr$^{-1}$.}
\tablenotetext{d}{Intrinsic Ly$\alpha$ luminosity measured in units of 10$^{43}$ ergs s$^{-1}$.}
\end{deluxetable}

\subsection{Interstellar Absorption}
\label{sec:abs}

The evidence for feedback in high-redshift galaxies comes in several forms, one of 
which is the nature of rest-frame UV interstellar absorption lines.  Interstellar 
material is swept up in a galaxy-wide outflow, causing a blueshift in the associated 
absorption lines.  In our sample over half of the objects have detected interstellar 
absorption lines.  These include both low-ionization (\ion{Si}{2}$\lambda$1260, 
\ion{Si}{2}$\lambda$1526, and \ion{C}{2}$\lambda$1334) and high-ionization 
(\ion{Si}{4}$\lambda\lambda$1393,1402 and \ion{C}{4}$\lambda\lambda$1548,1549) 
lines.

\begin{figure*}[htbp]
\begin{center}
\centerline{
%    \mbox{\includegraphics[scale =0.5]{abs_C49.eps}}
%    \mbox{\includegraphics[scale =0.5]{abs_bx154.eps}}
    \mbox{\includegraphics[scale =0.5]{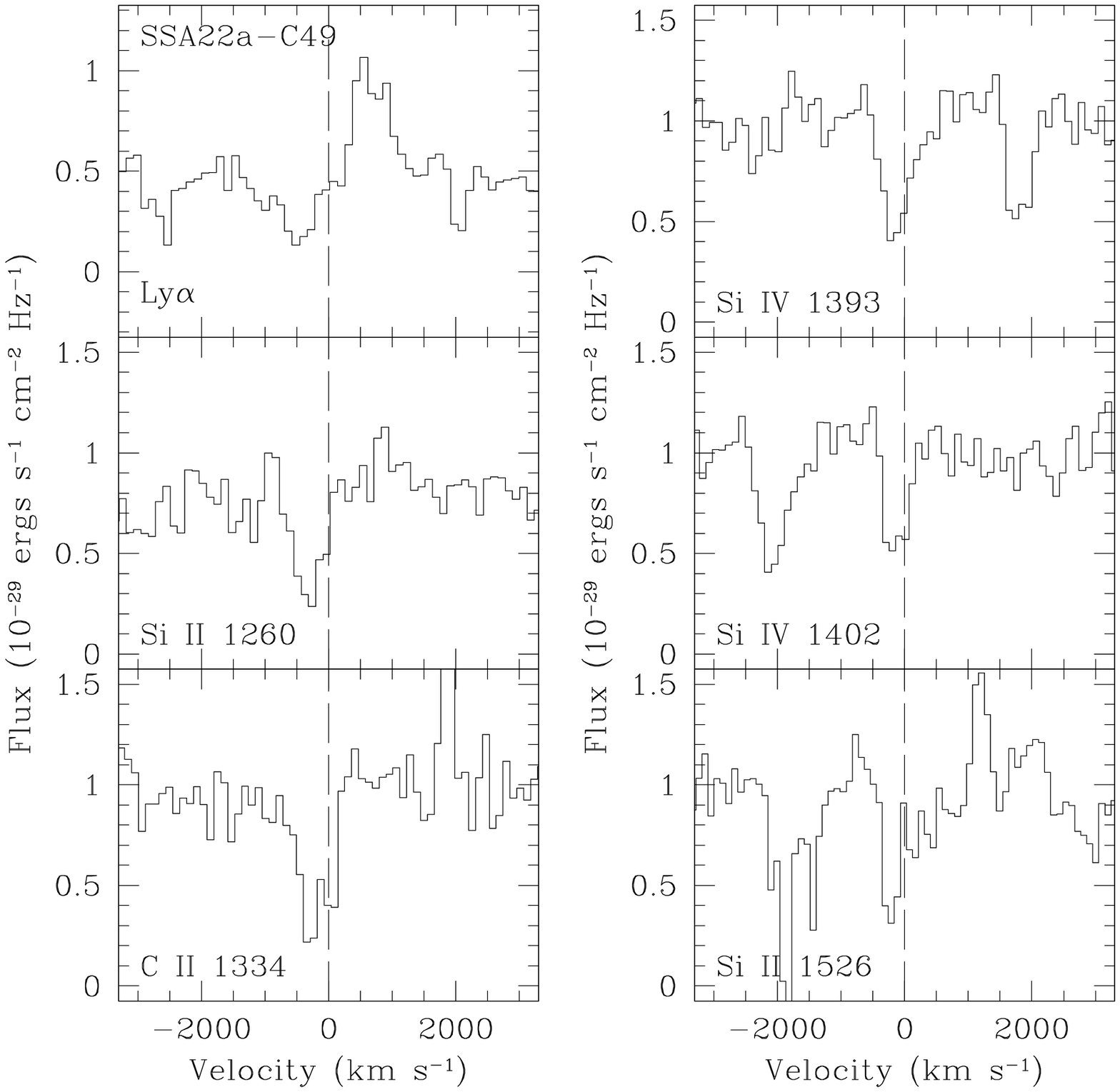}}
    \mbox{\includegraphics[scale =0.5]{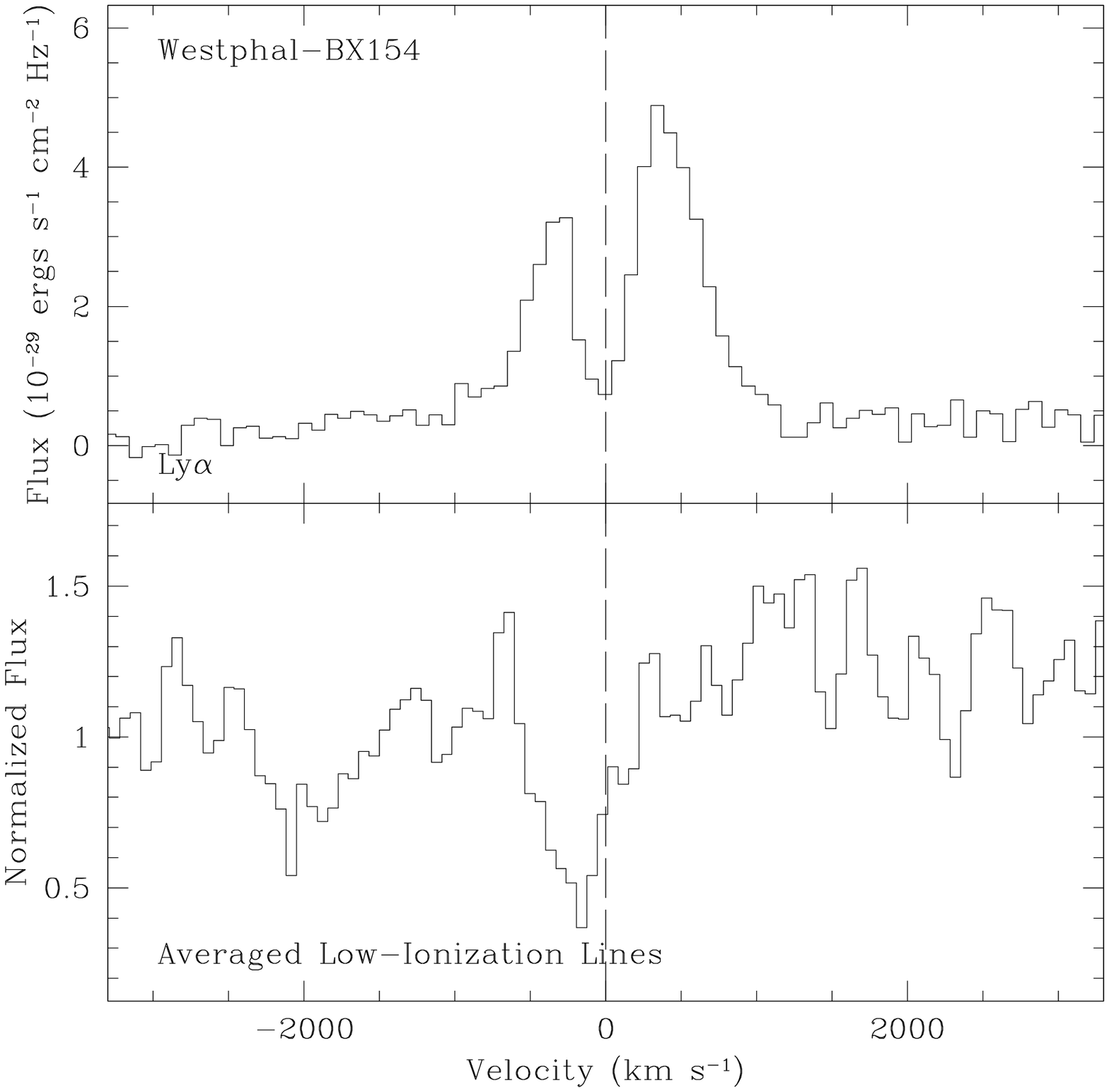}}
  }
\caption{Interstellar absorption velocity profiles for two objects in our sample. 
Interstellar absorption lines were measured for 12 out of the 18 objects in our sample, 
and the majority have centroids that are blueshifted relative to the velocity zeropoint.  (Left) 
Comparison of the Ly$\alpha$ and low- and high-ionization interstellar absorption velocity 
profiles for SSA22a-C49.  All of the absorption-line velocity centroids appear
blueward of the velocity-field zeropoint, whereas the 
Ly$\alpha$ velocity profile is shifted completely redward of the velocity-field zeropoint.  
(Right) Comparison of the Ly$\alpha$ and averaged low-ionization interstellar
absorption velocity profiles of Westphal-BX154. The low-ionization lines, 
\ion{Si}{2}$\lambda$1260, \ion{Si}{2}$\lambda$1526, and \ion{C}{2}$\lambda$1334, 
were averaged to obtain better S/N.  The absorption line velocity centroid is blueshifted
relative to the velocity-field zeropoint.  The two peaks of the Ly$\alpha$ velocity profile
straddle the velocity-field zeropoint.
\label{fig:absp}}
\end{center}
\end{figure*}

For all but two objects the absorption lines identified in the spectra are 
low-ionization species.  The S/N of these lines is not exceptionally strong on an 
individual basis.  In principle, all of the low-ionization lines arise in the same 
gas and should share the same $z_{abs}$ and velocity profiles.  Based on this idea 
we continuum-normalized and averaged the \ion{Si}{2}$\lambda$1260, 
\ion{Si}{2}$\lambda$1526, and \ion{C}{2}$\lambda$1334 velocity profiles to obtain 
better S/N.  Figure \ref{fig:absp} shows an example of one of our objects, 
Westphal-BX154, with averaged low-ionization lines, as well as object SSA22a-C49, 
which has measurements of both low- and high-ionization lines.  The velocity offsets 
in our sample measured from the absorption line centroids range from $\Delta 
v_{abs}=+14$ to $-1,174$ km s$^{-1}$, relative to the systemic velocity.  The 
average value of the interstellar absorption velocity offset for the sample is 
$\langle \Delta v_{abs} \rangle =-227$ km s$^{-1}$, which is indicative of 
large scale outflows \citep{ pettini01,shapley03,steidel10}.

\begin{figure}
\begin{center}
\centerline{
%    \mbox{\includegraphics[scale =0.80]{abs_sig.eps}}
    \mbox{\includegraphics[scale =0.45]{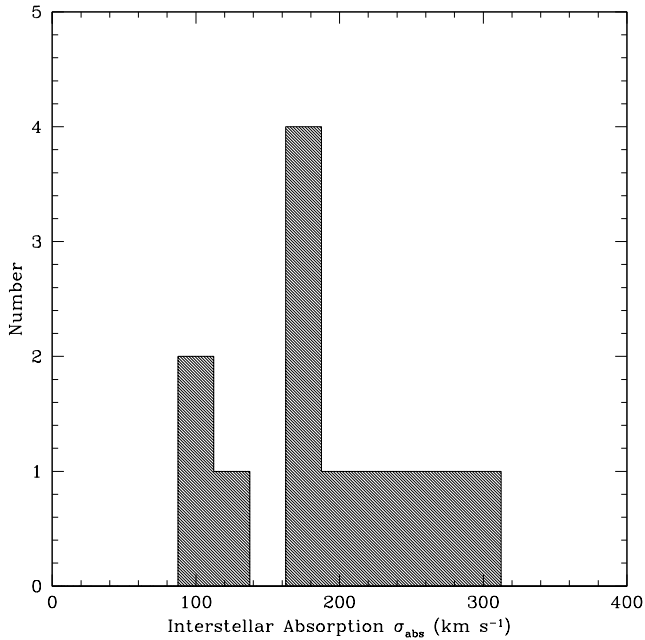}}
  }
\caption{Histogram of interstellar absorption line widths, $\sigma_{abs}$.  All 
$\sigma_{abs}$ values were measured from the combined low-ionization lines for each 
object, except for SSA22a-C32, SSA22a-C49 and SSA22a-D17, where individual 
low-ionization lines were measured and then averaged.  The mean value of the sample 
is $\langle \sigma_{abs} \rangle =$ 190 $\pm$ 63 km s$^{-1}$.  The line width is an important 
parameter for modeling Ly$\alpha$ radiative transfer because it reveals the dynamics 
of the absorbing interstellar gas.
\label{fig:abs_sig}}
\end{center}
\end{figure}

The line width of the interstellar absorption features was measured as 
$\sigma_{abs}=$ (FWHM/2.355) $\times(c/\lambda_{obs})$, where $\lambda_{obs}$ is the 
observed wavelength of the interstellar absorption line.  The FWHM was corrected for 
instrumental resolution ($\sigma_{inst}=$ (FWHM$_{inst}$/2.355) = 95, 155, 190 km 
s$^{-1}$ for the 600, 400, 300-line grism, respectively).  The line width, 
$\sigma_{abs}$, is a probe of the range of velocities present in the interstellar 
gas, which is an important observational constraint for Ly$\alpha$ radiative transfer models 
(see Section \ref{sec:model}).  The average $\sigma_{abs}$ for the 12 objects in our 
sample with absorption line measurements is 190 $\pm$ 63 km s$^{-1}$ (where
the error represents the standard deviation of the distribution of measurements).  This line 
width is consistent with measurements made by \citet{shapley03} of $\langle \sigma_{abs}\rangle$ = 
238 $\pm$ 64 km s$^{-1}$ in a composite spectrum of 811 $z\sim3$ Lyman Break 
Galaxies (LBGs) and \citet{quider09} of $\sigma_{abs} \sim$ 170$-$255 km s$^{-1}$ in 
the strong gravitationally lensed galaxy, ``the Cosmic Horseshoe", at $z=2.38$.  Our 
$\sigma_{abs}$ values were measured from the combined low-ionization lines for each 
object except for SSA22a-C32, SSA22a-C49 and SSA22a-D17, where individual 
low-ionization lines were measured and then averaged.  Figure \ref{fig:abs_sig} 
shows the distribution of the interstellar absorption line widths for our sample.

\begin{figure}[htbp]
\begin{center}
\centerline{
%    \mbox{\includegraphics[scale =0.80]{compare.eps}}
    \mbox{\includegraphics[scale =0.45]{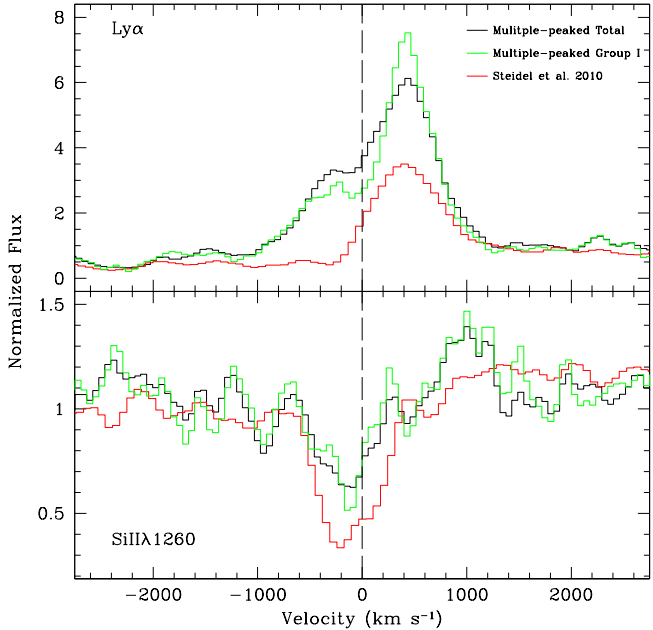}}
  }
\caption{Comparison of velocity profiles for the composite of the 18 multiple-peaked 
objects with NIRSPEC measurements, the subset of 11 objects that comprise Group I, 
and 29 single-peaked spectra from \citet{steidel10}.  (Top) The Ly$\alpha$ velocity 
profile for each composite spectrum.  The emission centroid from the single-peaked composite spectrum
matches the centroid of the red peak from both of the double-peaked composite 
spectra. (Bottom) The interstellar absorption velocity profile of 
\ion{Si}{2}$\lambda$1260 for each composite spectrum. The line widths for the 
single-peaked profile ($\langle \sigma_{abs} \rangle=220 \pm 19$ km s$^{-1}$) are systematically 
larger than for the multiple-peaked profile ($\langle \sigma_{abs} \rangle=140 \pm 26$ km s$^{-1}$).  
The line width for the Group I composite is $\langle \sigma_{abs} \rangle=100 \pm 31$ km s$^{-1}$.  
The interstellar absorption lines are also inherently deeper for the single-peaked 
compared to the multiple-peaked composite.
\label{fig:compare}}
\end{center}
\end{figure}

We also investigated if there was an inherent difference in the interstellar 
absorption line profiles for multiple-peaked compared to single-peaked Ly$\alpha$ 
emission spectra.  For this comparison we constructed composite LRIS rest-frame UV 
spectra for our sample of 18 multiple-peaked objects and for a sample of 29 galaxies 
from \citet{steidel10} with single-peaked Ly$\alpha$ emission profiles and precisely 
measured systemic redshifts based on NIRSPEC observations of H$\alpha$.  To create the composite spectrum we utilized the $\tt{IRAF}$ routine, $\tt{scombine}$,  computing the average of the rest-frame spectra at each wavelength with a minimum/maximum pixel rejection.  The composite spectrum was then continuum normalized using a cubic spline fitting function.  The Ly$\alpha$ profile in the composite spectrum of our sample is characterized by two clear peaks, with a stronger peak on the red side.  This profile is expected since 
the majority of our profiles resemble such a configuration (i.e. Group I).
For the single-peaked composite spectrum, the measured line widths of the interstellar 
absorption lines are systematically larger than those in the multiple-peaked 
composite spectrum, with $\langle \sigma_{abs} \rangle=140 \pm 26$ km s$^{-1}$ for the 
multiple-peaked composite and $\langle \sigma_{abs} \rangle=220 \pm 19$ km s$^{-1}$ for the 
single-peaked composite. The line widths, $\langle \sigma_{abs} \rangle$, were calculated from the averaged \ion{Si}{2}$\lambda$1260 and \ion{C}{2}$\lambda$1334 lines, which were chosen because of their higher S/N. A Monte Carlo approach, similar to the one described in Section \ref{sec:lineflux}, was used to estimate the error.  This difference in interstellar absorption linewidths may indicate that the range of velocities found within the interstellar gas must be narrower in order to create a multiple-peaked profile instead of a single-peaked profile.  A comparison of 
Ly$\alpha$ emission profiles indicates that the centroid of the Ly$\alpha$ emission 
peak in the single-peaked composite is the same as the centroid of the redder peak 
in the multiple-peaked composite spectrum.  Figure \ref{fig:compare} illustrates the 
differences between the total and Group I composite spectra 
from our sample and the stack of the 29 single-peaked 
Ly$\alpha$ emission objects from \citet{steidel10}.  We note here that only two
out of the 29 single-peaked spectra included in the single-peaked composite were taken
with the 600-line grism, while 600-line spectra comprise 5 out of 18 in the 
multiple-peaked composite.  To determine if a possible second peak was missed in the single-peaked composite due to decreased resolution, we smoothed the multiple-peaked composite to the resolution to the single-peaked composite. The second peak in the multiple-peaked 
composite is still distinctly evident after smoothing, demonstrating that the 
observed multiple-peaked phenomenon is not simply a result of better resolution.  Also of note is the fact that the redshift distribution for our multiple-peaked spectra is bimodal (with objects at $z\sim 2.1-2.6$ and $z\sim 3.0-3.3$) and characterized by a mean value of $\langle z \rangle =2.67 \pm 0.42$, with a 
median of $z=2.54$.  The redshift distribution for the single-peaked spectra 
has $\langle z \rangle =2.31 \pm 0.13$, with a median of $z=2.31$. To control for any possible effects of redshift evolution, we also compare the subset of $z\sim 2$ objects with the single-peaked spectra, and find the same spectral trends.

Finally, the interstellar absorption line kinematics allow us to consider the 
possibility of an alternate scenario for a multiple-peaked Ly$\alpha$ emission 
profile.  The multiple-peaked nature of Ly$\alpha$ emission can also be explained in 
terms of the interpeak trough tracing absorption from neutral HI with the same 
kinematics as the material giving rise to the interstellar metal absorption lines.  
According to this scenario, the trough and the interstellar absorption centroids 
would be aligned in velocity space.  Figure \ref{fig:vel2} shows that, with the 
possible exception of Q1549-BX202 and Q0449-BX167, the Ly$\alpha$ trough velocity 
centroid does not line up with the interstellar absorption centroid for any of the 
objects in our sample with interstellar absorption measurements -- ruling out this 
possible hypothesis.

\section{Ly$\alpha$ Radiative Transfer Modeling}
\label{sec:model}

\subsection{Models}
\label{sec:modeldescrip}

Explaining the emergent profile of Ly$\alpha$ emission in
star-forming galaxies has been the subject of many theoretical and
observational studies because of the easy observability of the
Ly$\alpha$ line and the complex processes that govern how it
emerges from galaxies. Models for the propagation and escape of
Ly$\alpha$ photons have been based on both Monte Carlo radiative
transfer and simple analytic calculations for a variety of
idealized gas density and velocity distributions
\citep{hansenoh06,steidel10,verhamme06,verhamme08} as well as
post-processing of 3D cosmological hydrodynamical simulations
\citep{laursen2009a,barnes2011}. For analytic models, the gas
density and velocity distributions assumed include static,
expanding, and infalling shells with both unity and non-unity covering fraction,
expanding and infalling uniform-density haloes in which the gas
velocity depends monotonically on galactocentric distance,
open-ended tubes, and spherical distributions of discrete clouds
with a range of velocity profiles. In some of these models, the
effect of dust extinction has been treated explicitly
\citep{verhamme06,verhamme08,hansenoh06, laursen2009b}.

One of the simple models that has been most closely compared with
individual LBG spectra is that of an expanding shell of gaseous
material \citep{verhamme06,verhamme08,schaerer2008}. In this
picture, energy from supernova explosions sweeps up the ISM of 
star-forming galaxies into a geometrically thin but optically thick
spherical shell. With the advent of efficient algorithms to perform
resonant-line radiative transfer, it has been shown that such a
system can be made to reproduce the observed Ly$\alpha$ profiles of
high-redshift galaxies \citep{verhamme06,verhamme08}.
\citet{verhamme06} deconstruct the Ly$\alpha$ line morphology in
order to determine how specific features in the Ly$\alpha$ profile
can be related to physical parameters of model systems as well as
specific photon trajectories. For example, a shell expanding with
uniform velocity $V_{\rm exp}$ will produce a primary redshifted
peak at approximately $v=2 V_{\rm exp}$ corresponding to photons
that have undergone a single backscattering event en route to the
observer.  Photons that undergo zero backscatterings emerge in an
asymmetric double-peaked structure centered about a velocity
$v=-V_{\rm exp}$. Finally, photons that experience multiple
backscattering events emerge redward of the primary peak.  At fixed
expansion velocity, the red-side peaks become indistinguishable
from one another and the blue-side peak becomes more blueshifted as the Doppler
parameter, $b$, increases. Increasing the column density but
keeping the expansion velocity and thermal velocity fixed results
in an enhanced inter-peak separation and broader peaks.
\citet{verhamme06} also consider the effect of dust attenuation on
the emergent Ly$\alpha$ profile, showing how it can reduce the
overall Ly$\alpha$ escape fraction, as well as modulate the shape
of the profile in velocity space.

\subsection{Comparison Between Models and Observations}
\label{sec:modelcomp}

\begin{figure}[htbp]
\begin{center}
\centerline{
%    \mbox{\includegraphics[scale =0.80]{juna_models.eps}}
    \mbox{\includegraphics[scale =0.53]{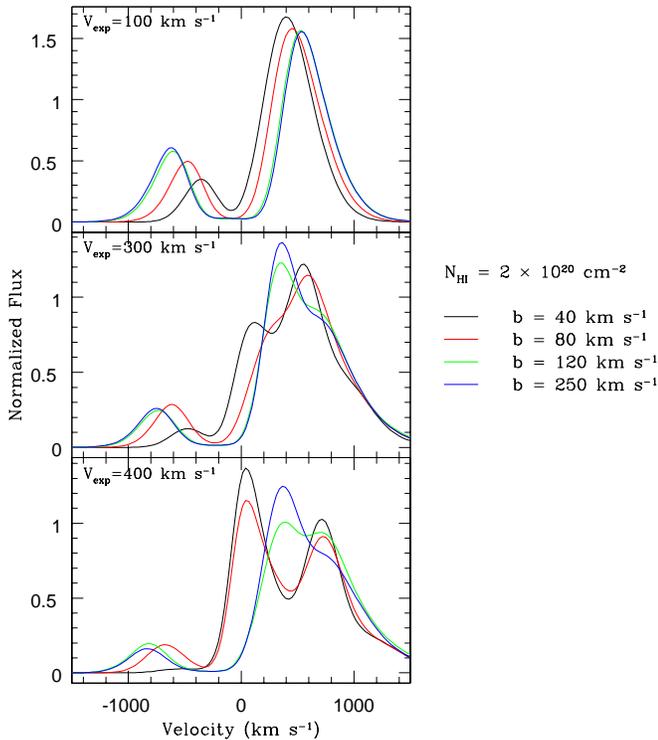}}
  }
\caption{Emergent Ly$\alpha$ profiles from an expanding shell
with a central monochromatic source. In all panels, the hydrogen
column density of the expanding shell is $N_{HI}=2\times 10^{20}\mbox{ cm}^{-2}$.
The top, middle, and bottom panels show the emergent profiles for shells
with expansion velocities of, respectively, $V_{\rm exp}=$100, 300, and 400
km~s$^{-1}$. In each panel, black, red, green, and blue curves respectively indicate
profiles for Doppler parameters of $b=$40, 80, 120, and 250 km~s$^{-1}$.
Each predicted model curve has been smoothed to the typical resolution of our
LRIS spectra.
\label{fig:kollmeier_models}}
\end{center}
\end{figure}

In order to investigate whether such a system can reproduce the
data presented here, we use the radiative transfer code developed
in \citet{zheng02} and \citet{kollmeier2010} to generate the
observed Ly$\alpha$ spectra emerging from expanding/infalling
shell models similar to those presented in \citet{verhamme06},
with a range of shell column densities, expansion velocities and
Doppler parameters.
Figure~\ref{fig:kollmeier_models} shows a subset of such models,
for a central monochromatic source of Ly$\alpha$ photons propagating
outward through an expanding shell with a column density of $N_{HI}=2
\times 10^{20} \mbox{ cm}^{-2}$, expansion velocities of $V_{\rm exp} =$
100, 300, and 400 km s$^{-1}$, and Doppler $b$-parameters
of 40, 80, 120, and 250 km~s$^{-1}$.  It should be noted that the Ly$\alpha$ radiative transfer calculation does not depend on the exact physical dimensions of the system for a given column density, expansion velocity, and Doppler parameter as long as the shell is kept geometrically thin.  The models have been smoothed
to the typical resolution of our LRIS data, for a fair comparison
with the observations. Future modeling (Kollmeier et al., in prep.)
will include a distribution of injection frequencies for Ly$\alpha$
photons that matches the nebular velocity dispersions measured from
H$\alpha$ and [OIII], as well as a more extensive set of gas
density and velocity distributions for radiative transfer
calculations.

\begin{figure}[htbp]
\begin{center}
\centerline{
%    \mbox{\includegraphics[scale =0.80]{models_group1.eps}}
    \mbox{\includegraphics[scale =0.53]{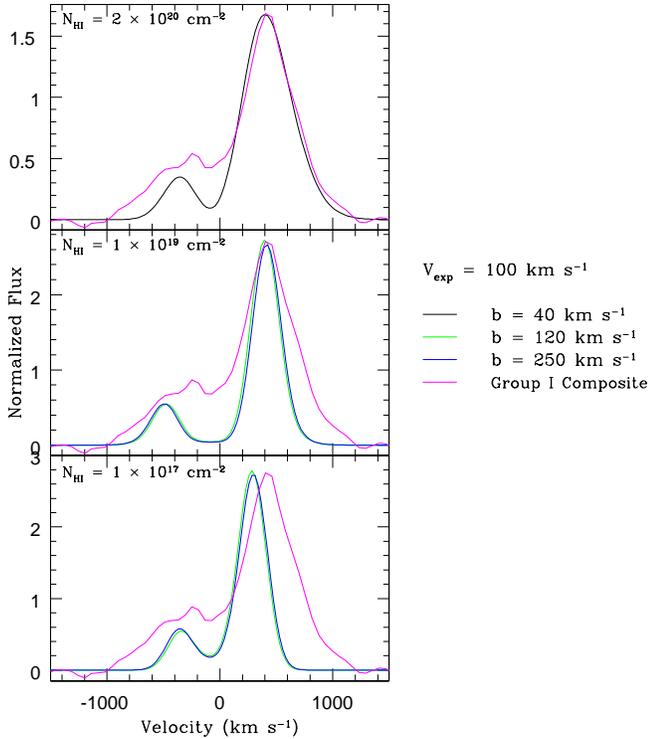}}
  }
\caption{Emergent Ly$\alpha$ profiles from simple shell models
that are qualitatively similar to Group I spectra. All panels
show the predicted profiles for a central monochromatic
source passing through an expanding shell with $V_{\rm exp}=$100 km~s$^{-1}$.  The top panel features a model with $N_{HI}=2\times 10^{20}\mbox{ cm}^{-2}$and Doppler parameter $b=40$~km~s$^{-1}$. The middle and bottom panels show the results for models with $N_{HI}=1\times 10^{19}\mbox{ cm}^{-2}$
and $N_{HI}=1\times 10^{17}\mbox{ cm}^{-2}$, respectively. In both
of these panels, the predicted profiles are shown for Doppler parameters
$b=$~120 and 250 km~s$^{-1}$. In all panels, a double-peak profile is
shown, with two peaks straddling the velocity field zeropoint, and a stronger
red peak. Overplotted on each panel is the Group I continuum-subtracted composite spectrum.  The Group I spectrum was normalized to match the primary peak height of each model.  The $N_{HI}=2\times 10^{20}\mbox{ cm}^{-2}$ model has the closest resemblance to the typical properties of the Group I Ly$\alpha$ spectra in terms of both precise peak locations and line widths.  The $N_{HI}=1\times 10^{17}\mbox{ cm}^{-2}$ model displays peaks that are both too close together in velocity space and too narrow.
While the peak locations in the $N_{HI}=1\times 10^{19}\mbox{ cm}^{-2}$ model are a fine match
to the Group I composite spectrum, the predicted line widths are still too narrow.
\label{fig:groupI_models}}
\end{center}
\end{figure}

By far the most common Ly$\alpha$ profile observed within the
NIRSPEC multiple-peaked sample is the ``Group I" type, with two
peaks straddling the velocity-field zeropoint, and a stronger
red-side peak. Given the frequency of this profile type, it is of
particular interest to compare it with Ly$\alpha$ radiative
transfer models. In addition to the Ly$\alpha$ emission properties,
we mention here that the typical low-ionization interstellar
absorption blueshift for Group I objects is slightly smaller than
the sample average, with $\langle \Delta v_{abs}
\rangle_{\mbox{Group I}} = -90$~km~s$^{-1}$ (based on the 7 out of
11 Group I objects with interstellar absorption line measurements).
The average low-ionization interstellar absorption line width for
Group I objects is $\langle \sigma_{abs} \rangle \sim 160$~km~s$^{-1}$,
corresponding to $b=226$~km~s$^{-1}$, if the line width is
interpreted as the thermal velocity of a single expanding component
of gas \citep[which may indeed not be the correct interpretation of
the interstellar absorption profiles, see, e.g., ][and Section~\ref{sec:caveats}]{steidel10}.
As seen in Figure~\ref{fig:kollmeier_models}, the best match with
the typical Group I profile is obtained using the $N_{HI}=2
\times 10^{20} \mbox{ cm}^{-2}$, $V_{\rm
exp}=100$~km~s$^{-1}$, $b=40$~km~s$^{-1}$ model. In this model,
which also features two peaks straddling the velocity-field
zeropoint and a stronger red peak, the blue peak appears at $\Delta
v = -370$~km~s$^{-1}$, while the red peak appears at $\Delta v =
+430$~km~s$^{-1}$. The model matches the average Group I properties
both in terms of the locations of the Ly$\alpha$ peaks relative to
zero, as well as the interpeak separation, although the predicted contrast
between the red and blue peak heights is too large. At the same time, the predicted
absorption profile for such an expanding shell (which should be
traced well by the low-ionization features arising in neutral hydrogen
gas) would have $\langle
\sigma_{abs} \rangle \sim b / \sqrt{2} \sim 30$~km~s$^{-1}$, i.e., {\it significantly} lower
than what is observed for Group I objects. This difference
represents a fundamental discrepancy between the shell model and
the full complement of Group I data -- even if the average Ly$\alpha$
emission profile can be roughly reproduced. 
Figure \ref{fig:groupI_models} shows the $N_{HI}=2
\times 10^{20} \mbox{ cm}^{-2}$, $V_{\rm
exp}=100$~km~s$^{-1}$, $b=40$~km~s$^{-1}$ model, along with other
$V_{\rm exp}=100$~km~s$^{-1}$ expanding shell models with
smaller column density (i.e., $N_{HI}=1 \times 10^{17} \mbox{ cm}^{-2}$ or
$N_{HI}=1 \times 10^{19} \mbox{ cm}^{-2}$) and larger Doppler parameters ($b\geq 120 $~km~s$^{-1}$).  Overplotted on each panel is the continuum-subtracted composite spectrum from Group I, which has been normalized to the primary peak height of each model.  The models with the smaller column density also predict
emergent profiles with two peaks straddling the velocity zeropoint
and a stronger red-side peak.  However, in comparison to the Group I composite spectrum, the smallest
column density model features a peak velocity separation
that is too small ($\Delta v=640 $~km~s$^{-1}$), and, in both lower column density cases,
the emission peaks themselves are too narrow.  An important limitation in this analysis is the oversimplification of comparing a model characterized by a specific set of physical parameters with a composite spectrum. While the composite spectrum offers a boost in S/N relative to an individual spectrum, it also includes the spectra of many individual galaxies that likely span a range of parameters (e.g. $N_{HI}, b, V_{\rm exp}$).
 
The Ly$\alpha$ profiles for objects in Group II are also
characterized by two peaks straddling the velocity-field
zeropoint, but with a stronger blue peak. A double-peaked profile
with a dominant blue peak represents a key signature of infalling
gas \citep{zheng02,dijkstra2006a,verhamme06,barnes2011}. In terms of the
shell models presented in Figure~\ref{fig:kollmeier_models}, the
predicted Ly$\alpha$ profile for an infalling shell of material
with a specific column density and Doppler parameter can be
derived by simply flipping the corresponding expanding-shell model
about the line $\Delta v=0$. As shown in \citet{verhamme06} and
\citet{dijkstra2006a}, infalling spherical haloes of gas will also
yield double-peaked profiles with a stronger blue peak, assuming
either uniform emissivity or a central point source. The lack of
significant blueshift in the observed low-ionization interstellar
absorption lines of Group II objects (with Q1549-C20 actually
showing a slight redshift) lends additional support to a model of
infalling gas. In detail, however, there are mismatches between
the particular infall models mentioned above and the observed
Group II spectra. The average peak separation for the Group II
objects is $\langle \Delta v_{peak} \rangle \sim 800$~km~s$^{-1}$,
with the blueshifted peak located at $\Delta v \sim
-200$~km~s$^{-1}$. The infalling halo model of \citet{verhamme06}
predicts a peak separation of only $\langle \Delta v_{peak}
\rangle \sim 460$~km~s$^{-1}$, which is significantly smaller than
the observed one (see their Figure 5). While an infalling shell model with $V_{\rm
infall}=100$~km~s$^{-1}$ and $b= 40 $~km~s$^{-1}$ produces roughly
the correct peak separation and stronger blue peak, the locations
of the blue and red peaks in the infalling shell model relative to
zero velocity do not match those of the observed Group II spectra.

Like the spectra in Group II, Group III spectra also exhibit a stronger
blue peak in the Ly$\alpha$ profile, but both blue and red peaks are
redshifted relative to the velocity-field zeropoint, and most likely trace
outflowing rather than inflowing gas. Only one of the two Group III
objects (SSA22a-C49) has measurable low-ionization interstellar
absorption.  From these features, we measure a significant blueshift of
$\Delta v \sim -230$~km~s$^{-1}$, and a line width of $\sigma_{abs} \sim
230$~km~s$^{-1}$. In terms of the simple shell models shown in
Figure~\ref{fig:kollmeier_models}, the closest match to Group III is found
for $V_{\rm exp}=400$~km~s$^{-1}$, and $b= 40 $~km~s$^{-1}$. As discussed
by \citet{yang2011}, the blueshifted peak in a shell model decreases in
strength as $V_{\rm exp}/b$ increases. This trend is reflected in the
large $V_{\rm exp}/b$ model discussed here, with its
lack of significant flux bluewards of zero, and
two peaks redshifted relative to zero. Although there is qualitative
agreement between the model and Group III spectra, in so far as both have
two peaks redshifted relative to zero, and no significant flux bluewards of
zero, the actual peak separations and locations in the model and data do
not agree in detail. The model predicts $\Delta v_{peak} \sim
600$~km~s$^{-1}$, while the Group III objects have $\Delta v_{peak} \sim
400$~km~s$^{-1}$. Furthermore, the blue peak in the model is shifted by
only $\sim +100$~km~s$^{-1}$ relative to zero, while the blue peak in
SSA22a-C49 is shifted by $>+500$~km~s$^{-1}$ relative to zero. Finally,
the relatively small Doppler parameter of $b= 40 $~km~s$^{-1}$ would give
rise to significantly narrower low-ionization interstellar absorption
line widths than those observed. While Group III objects most likely
indicate the presence of an outflow, the models described here still present some mismatches with the Group III data.

Distinguishing between Group II and Group III objects by
establishing the velocity-field zeropoint has potentially
important consequences. Over the last several years, 
the process of galaxy growth through cold gas accretion has received much
attention in the theoretical literature. Both
analytic calculations \citep{birnboimdekel2003} and numerical simulations
\citep{fardal01,keres2005,keres2009,dekel2009} suggest that high-redshift galaxies
primarily grow by smoothly accreting cold gas from
the surrounding IGM.  Identifying observational signatures of
infalling gas is crucial for testing the theoretical paradigm of
cold accretion.  In particular,
since the predicted signatures of accreting gas in low-ionization
metal absorption lines are so weak, due to the low metallicity of
infalling material \citep{fumagalli2011}, the Ly$\alpha$ feature
may represent the best hope for detecting cosmological infall.
Beyond the Group II and Group III objects in our NIRSPEC sample,
$\sim 30$\% of the double-peaked spectra in the LRIS parent sample
have a stronger blue peak. If the velocity-field zeropoint can be
established in these objects, to test whether the stronger blue
peak is also blueshifted relative to zero (i.e., distinguishing
between Group II and Group III spectra), their Ly$\alpha$ spectra
may provide evidence for and constraints on the nature of gas infall in
star-forming galaxies at $z\sim 2-3$.

For the remaining three spectra (Groups IV and V), we do not find
even qualitative matches among the simple models presented here.
Indeed, for SSA22a-D17 and SSA22a-C31 (Group IV), there are two
peaks with roughly equal heights, as seen in a static
configuration. However, the locations of the peaks are not
symmetrically distributed around the velocity-field zeropoint, as
would be expected for a static shell, and the interpeak trough is
redshifted by several hundred km~s$^{-1}$. Furthermore, SSA22a-D17
has an extremely large low-ionization interstellar absorption
blueshift measured, with $\Delta v \sim -1170$~km~s$^{-1}$, which
additionally rules out the static case. Q1623-BX129 has two strong
peaks with roughly equal heights and similar locations to those of
SSA22a-D17 and SSA22a-C31, plus a third, redder peak, which is
significantly weaker. A larger model parameter space will be required to
match the profile shapes of these three objects.

In summary, while general qualitative matches can be found for the
Group I, Group II, and Group III Ly$\alpha$ profile types, we find
notable specific discrepancies between the models and the totality
of the data presented here -- with particular attention to the
interstellar absorption profiles. One important issue is that the
shell models that provide the best matches to, e.g., Group I
profiles ($V_{\rm exp}=100$~km~s$^{-1}$, $b=40$~km~s$^{-1}$), are
not able to match simultaneously the broad low-ionization
interstellar absorption troughs characterized by a median of
$\sigma_{abs,med}\sim 190$~km~s$^{-1}$. We also highlight the
importance of our new NIRSPEC measurements for untangling the
nature of these Ly$\alpha$ spectra. Measurements of the
velocity-field zeropoint (i.e. from the H$\alpha$ or [OIII]
redshift) are key for distinguishing between Group II (inflow) and
Group III (strong outflow) objects. Furthermore, determining the
velocity width of the input Ly$\alpha$ spectrum (based on the
H$\alpha$ or [OIII] velocity dispersion), so that it is not a free
parameter of Ly$\alpha$ radiative transfer models, reduces the
degeneracies associated with fitting physical models to Ly$\alpha$
profiles. Indeed, when fitting the double-peaked Ly$\alpha$
profile for an object with unconstrained systemic velocity and
input Ly$\alpha$ velocity dispersion, \citet{verhamme08} find two
expanding-shell model solutions that differ by 5 orders of
magnitude in inferred hydrogen column density (see their Figure 8). One of the best-fit
models has an inferred intrinsic Ly$\alpha$ FWHM$=700$~km~s$^{-1}$
-- a factor of three larger than the average for our sample!
Furthermore, the placement of the velocity-field zeropoint differs
by $\sim 500$~km~s$^{-1}$ for the two best-fit models in question.  \citet{verhamme08} conclude that neither solution is satisfactory and that future observations are needed to constrain their models. 
With systemic redshift and velocity dispersion measurements, such
degeneracy in best-fit models is no longer allowed, potentially
leading to much tighter constraints on the underlying physical
picture.

\subsection{Additional Considerations and Caveats}
\label{sec:caveats}

In the preceding discussion, within the framework of expanding/infalling
shell models, we presented an interpretation of the observed low-ionization
interstellar absorption line widths in terms of the random thermal         
broadening of gas in the shell. An alternative explanation of interstellar
absorption line widths is presented in \citet{steidel10}, according to
which absorption is produced by a population of discreet clouds at a large
range of galactocentric radii (i.e., not a thin shell). The expansion           
velocities of these clouds increase smoothly and monotonically with increasing radius, and
their covering fraction is a decreasing function of radius.
In such a model, in which photons scatter off of the $\tau_{Ly\alpha}=1$
surfaces of discrete clouds, the nature of the bulk gas kinematics modulates the emergent
Ly$\alpha$ profiles, as opposed to the neutral hydrogen column density
and Doppler $b$-parameter of individual clouds. 
This clumpy outflow model accounts for the typical interstellar absorption profiles
both along the line of sight to individual UV-selected star-forming     
galaxies at $z\sim 2-3$, as well as along averaged offset sightlines with
impact parameters up to $\sim 100$ physical kpc, and predicts
extended Ly$\alpha$ emission consistent with the observations \citep{steidel11}.   At the same time, high-resolution 3D
cosmological hydrodynamical simulations that follow the radiative transfer
of Ly$\alpha$ photons through the gas distributions surrounding high-redshift
galaxies are highlighting the importance of orientation effects.
The gas column density and velocity distributions in these simulated
galaxies are anisotropic and irregular, characterized by clumps and
filaments of infalling and outflowing cold gas \citep{laursen2009a,barnes2011}.
This anisotropy leads to a strong dependence of the shape of the emergent
Ly$\alpha$ profile on viewing angle. For example, if the line of sight to a galaxy
intercepts a filament of infalling gas, a blue-asymmetric peak may be
observed, even in the presence of outflowing gas in the system
\citep{barnes2011}.  A close comparison is
required between the predictions of such simulations and the full ensemble
of Ly$\alpha$ profiles observed in LBGs, as well as incorporating detailed Ly$\alpha$ radiative transfer calculations from \citet{zheng02} and
\citet{kollmeier2010} into the clumpy outflow model of \citet{steidel10}.

Another potentially important effect to consider is absorption by the IGM \citep{dijkstra2007,zheng10}.
Recent simulations by \citet{laursen2011} have demonstrated that the blue peak in
an emergent double-peaked Ly$\alpha$ profile may be suppressed due to absorption by neutral
hydrogen external to a galaxy. For the purposes of such analysis, ``IGM" is 
defined as gas located at a radius greater than $\sim 1.5$ times the
dark-matter halo virial radius. IGM absorption may be significant even
at redshifts well below the epoch of reionization. This result is relevant
to the interpretation of our Group I profile types. In such objects,
the Ly$\alpha$ peak bluewards of the systemic velocity is weaker than
the red peak. We have compared these profiles to simple galactic-scale models for outflowing
gas, in which the asymmetry arises due to the radiative transfer of Ly$\alpha$ photons,
and have ignored the possible effects of IGM absorption on the observed
red-to-blue peak ratio. One simple test for the importance of IGM absorption
consists of examining the distribution of red-to-blue peak ratios as a function
of redshift within the LRIS double-peaked parent sample. For this test, we
divide the sample into high- ($z\sim 3$) and low-redshift ($z\sim 2$)
subsamples at a threshold redshift of $z=2.7$, and calculate the 
median red-to-blue peak ratio for each subsample. If IGM effects
are significant, we expect to see the red-to-blue peak ratio 
evolve towards lower values at lower redshifts, as the blue peak becomes
less suppressed by IGM absorption (under the assumption that the two
samples have the same intrinsic distributions of double-peaked Ly$\alpha$ profile shapes 
before being subject to IGM absorption). In fact, we find that
the median red-to-blue peak flux ratios for $z\sim3$ 
and $z\sim 2$ subsamples are 1.3 and 1.5, respectively, and not statistically
distinguishable. Therefore, IGM absorption does not appear to have a significant
impact on the observed multiple-peaked Ly$\alpha$ profile shapes in our sample.

Finally, we must admit the possibility of an entirely different
explanation for the origin of multiple-peaked Ly$\alpha$ emission
profiles, such as galaxy mergers. A multiple-peaked profile would naturally
result from a system in a merger state, with different peaks corresponding
to different merging components \citep{cooke10,rauch11}. The continuum and
emission-line morphology of objects in our sample are crucial to consider
when examining possible merger models. Figure \ref{fig:spec1} shows the
(seeing-limited) $\cal R$-band morphologies for objects in our sample.
The majority of our targets appear to be consistent with single components
at this resolution.  A small number of objects, however, are slightly
extended (e.g. Q2206-BX151, Q1549-M22) or appear to have multiple
components (e.g. Q0449-BX167, Westphal-BX154).  Three of these objects
have already been discussed in Section \ref{sec:resolved} in terms of
their complex two-dimensional spectra.  If our multiple-peaked Ly$\alpha$
profiles were simply caused by mergers, there should be evidence of
multiple-peaked emission in the rest-frame optical nebular lines (i.e.,
H$\alpha$ and [OIII]).  These features more closely trace the systemic
velocity and dynamics and are typically not as sensitive to bulk outflow
motions as the Ly$\alpha$ emission line. As described in Section
\ref{sec:resolved}, only Q1549-M22 shows possible signs of nebular lines
with multiple-peaked behavior in the spectral direction, ruling out
mergers as a possible explanation for the majority of our Ly$\alpha$
profiles.

\section{Summary and Discussion}
\label{sec:conclusions}

There is strong observational evidence for the ubiquity of large-scale
outflows in high-redshift star-forming galaxies. Gauging the overall impact
of these large-scale outflows on the evolution of the galaxies sustaining
them remains an open challenge. Theoretical considerations suggest that,
at the same cosmic epochs, the {\it inflow} of cold gas is also a very important
process in the evolution of star-forming galaxies. Multiple-peaked Ly$\alpha$ 
profiles in star-forming UV-selected galaxies at $z\sim2-3$ offer a unique probe of 
both outflows and inflows in the early universe.  
The complex line structure potentially provides 
constraints on many important gas flow parameters.  Specifically, the locations of 
peaks with respect to the velocity-field zeropoint, their separations, and flux 
ratios are all additional pieces of information not seen in single-peaked Ly$\alpha$ 
profiles.  Along with observational data such as accurate systemic redshifts, 
intrinsic nebular line widths, and intrinsic ionizing photon fluxes, one can 
distinguish among the possible processes underlying these systems.

We have shown that the phenomenon of multiple-peaked profiles appears in a 
significant fraction ($\sim$ 30$\%$) of UV-selected star-forming galaxies at 
$z\sim2-3$ with Ly$\alpha$ emission.  After identifying possible candidates for this 
study we presented a sample of 18 multiple-peaked objects with observed H$\alpha$ or 
[OIII] nebular emission lines, which were used to establish accurate systemic 
redshift measurements.  The average velocity dispersion, $\sigma_{v}$, and the mean 
star-formation rate in our sample are comparable to values measured from the full 
population of UV-selected star-forming galaxies at $z\sim2-3$.  At the same time, 
given our focus on Ly$\alpha$-emitting galaxies, the average $E(B-V)$ in our sample 
is bluer than what is typically measured.  Detected interstellar absorption lines 
for objects in the sample show, on average, a blueshift, which is suggestive of 
large-scale outflows driven by supernovae or winds of massive stars.  Additionally, 
the interstellar absorption line widths are measured to be $\langle \sigma_{abs} \rangle\sim 
200$ km s$^{-1}$, which is similar to values found in other studies of star-forming 
galaxies at $z\sim2-3$.  A comparison of our interstellar absorption line widths to 
a sample of star-forming galaxies with single-peaked Ly$\alpha$ emission show the 
single-peaked objects to have consistently larger interstellar absorption line 
widths.  The difference in interstellar absorption line widths may signify that the 
ranges of gas velocities are different for multiple-peaked compared to single-peaked 
Ly$\alpha$ emission objects.

We have qualitatively compared our observed Ly$\alpha$ profiles with
simple models of expanding and infalling gaseous shells and halos, given
the attention such models have recently received in the literature \citep{verhamme06,
verhamme08,schaerer2008}. The closest match between these simple models and our data
is found for Group I type profiles, which make up 11 out of 18 
of the NIRSPEC sample objects, and are characterized by a red-asymmetric double peak
straddling the velocity-field zeropoint.
Group I type profiles can be produced in an expanding
shell model, with a high-column density ($N=2 \times 10^{20} \mbox{ cm}^{-2}$), moderate 
expansion speed ($V_{\rm exp}=100$~km~s$^{-1}$), and small Doppler 
parameter, ($b=40$~km~s$^{-1}$). At the same time, the observed low-ionization
interstellar absorption line profiles for Group I objects are significantly
broader than the features that would arise from an expanding shell with
$b=40$~km~s$^{-1}$, and signal a problem with the above interpretation
for the Group I Ly$\alpha$ spectra. Certain aspects of Group II and Group III
Ly$\alpha$ spectra are also reproduced by the simple models considered here,
though, in detail, the matches are not perfect. For the remaining handful
of objects in our NIRSPEC multiple-peaked sample, the suite of models
we've considered cannot be used to explain the observed profiles, both
in terms of Ly$\alpha$ peak locations, relative strengths, and numbers.

It is also of interest to compare Ly$\alpha$ radiative transfer models
with high-redshift data of significantly higher S/N and resolution than the 
spectra presented here.  \citet{quider09} present the multiple-peaked 
Ly$\alpha$ emission profile of the strongly gravitationally lensed 
galaxy, ``the Cosmic Horseshoe," for which the magnification is estimated
to be a factor of $\sim 24$.  The Cosmic Horseshoe has a very well-defined
stellar systemic redshift, from which the Ly$\alpha$ spectral profile can be calculated in
velocity space. While an expanding shell model from \citet{verhamme06}
with $N_{HI}=7 \times 10^{19} \mbox{ cm}^{-2}$, $V_{\rm exp}=300$~km~s$^{-1}$,
and $b=40$~km~s$^{-1}$, provides a qualitative match to the two redshifted
peaks in the Cosmic Horseshoe Ly$\alpha$ profile, the small
Doppler $b$-parameter in the model is again at odds with the broad low-ionization interstellar
absorption profile. Furthermore, in detail, the locations of the Ly$\alpha$ peaks
in the predicted model spectrum do not line up with those of the observations.
Both typical and strongly-lensed UV-selected star-forming galaxies with well-determined
systemic redshifts and, accordingly, Ly$\alpha$ profiles calculated robustly
in velocity space are now providing important observational constraints on Ly$\alpha$ radiative
transfer models.  These objects highlight the need for considering simultaneously
the emergent Ly$\alpha$ emission profile, and the profile of
low-ionization interstellar absorption, which must be matched
within a unified framework \citep{steidel10}.

While we have mainly focused here on the discrepancies between models and data,
and have therefore not yet been able to derive the physical parameters
of the gaseous flows surrounding UV-selected star-forming galaxies at high redshift,
we view this analysis as an important initial step. It will simply not be possible
to constrain the gas density and velocity distributions in these circumgalactic
flows without the simultaneous
establishment of the location of the multiple-peaked Ly$\alpha$ profile and interstellar
absorption features in velocity space. We have presented such measurements here for 
a sample of objects whose red- and blue-asymmetric multiple-peaked Ly$\alpha$ profiles 
at least qualitatively suggest a range of processes including both outflows and infall. 
Additional critical observables presented here which will be be considered in future modeling
are the input velocity distribution of Ly$\alpha$ photons as traced
by the velocity dispersion of either H$\alpha$ or [OIII] emission, and the 
degree of dust attenuation as traced by rest-frame UV colors and Ly$\alpha$/H$\alpha$
flux ratios.  With a realistic model that successfully 
matches the combination of intrinsic and emergent 
Ly$\alpha$ velocity distributions as well as interstellar absorption profiles,
we will obtain the gas density and velocity distribution as a function of
radius. These distributions will potentially allow us to constrain the rates at which
cold gas mass is flowing in and out of galaxies, which represents one of the 
most important goals in the study of galaxy formation.

\acknowledgments 
We thank Mark Dijkstra for helpful discussions that enhanced the presentation
of our results.  K.R.K. and A.E.S. acknowledge support from the 
David and Lucile Packard Foundation. C.C.S. acknowledges 
additional support from the John D. and Catherine T. MacArthur Foundation, 
the Peter and Patricia Gruber Foundation, and NSF grants AST-0606912 and AST-0908805. 
Z.Z. gratefully acknowledges support from 
Yale Center for Astronomy and Astrophysics through a YCAA fellowship. 
We wish to extend special thanks to those of Hawaiian ancestry on whose 
sacred mountain we are privileged to be guests. 
Without their generous hospitality, most of the observations presented 
wherein would not have been possible.

\bibliographystyle{apj}
\bibliography{doublepeak}

\begin{thebibliography}{53}
\expandafter\ifx\csname natexlab\endcsname\relax\def\natexlab#1{#1}\fi

\bibitem[{{Adelberger} {et~al.}(2005){Adelberger}, {Steidel}, {Pettini},
  {Shapley}, {Reddy}, \& {Erb}}]{adelberger05}
{Adelberger}, K.~L., {Steidel}, C.~C., {Pettini}, M., {Shapley}, A.~E.,
  {Reddy}, N.~A., \& {Erb}, D.~K. 2005, \apj, 619, 697

\bibitem[{{Adelberger} {et~al.}(2003){Adelberger}, {Steidel}, {Shapley}, \&
  {Pettini}}]{adelberger03}
{Adelberger}, K.~L., {Steidel}, C.~C., {Shapley}, A.~E., \& {Pettini}, M. 2003,
  \apj, 584, 45

\bibitem[{{Ahn} {et~al.}(2002){Ahn}, {Lee}, \& {Lee}}]{ahn02}
{Ahn}, S., {Lee}, H., \& {Lee}, H.~M. 2002, \apj, 567, 922

\bibitem[{{Barnes} {et~al.}(2011){Barnes}, {Haehnelt}, {Tescari}, \&
  {Viel}}]{barnes2011}
{Barnes}, L.~A., {Haehnelt}, M.~G., {Tescari}, E., \& {Viel}, M. 2011, ArXiv
  e-prints

\bibitem[{{Birnboim} \& {Dekel}(2003)}]{birnboimdekel2003}
{Birnboim}, Y. \& {Dekel}, A. 2003, \mnras, 345, 349

\bibitem[{{Bruzual} \& {Charlot}(2003)}]{bruzchar03}
{Bruzual}, G. \& {Charlot}, S. 2003, \mnras, 344, 1000

\bibitem[{{Calzetti} {et~al.}(2000){Calzetti}, {Armus}, {Bohlin}, {Kinney},
  {Koornneef}, \& {Storchi-Bergmann}}]{calzetti2000}
{Calzetti}, D., {Armus}, L., {Bohlin}, R.~C., {Kinney}, A.~L., {Koornneef}, J.,
  \& {Storchi-Bergmann}, T. 2000, \apj, 533, 682

\bibitem[{{Chabrier}(2003)}]{chabrier03}
{Chabrier}, G. 2003, \pasp, 115, 763

\bibitem[{{Cooke} {et~al.}(2010){Cooke}, {Berrier}, {Barton}, {Bullock}, \&
  {Wolfe}}]{cooke10}
{Cooke}, J., {Berrier}, J.~C., {Barton}, E.~J., {Bullock}, J.~S., \& {Wolfe},
  A.~M. 2010, \mnras, 403, 1020

\bibitem[{{Dekel} {et~al.}(2009){Dekel}, {Birnboim}, {Engel}, {Freundlich},
  {Goerdt}, {Mumcuoglu}, {Neistein}, {Pichon}, {Teyssier}, \&
  {Zinger}}]{dekel2009}
{Dekel}, A., {Birnboim}, Y., {Engel}, G., {Freundlich}, J., {Goerdt}, T.,
  {Mumcuoglu}, M., {Neistein}, E., {Pichon}, C., {Teyssier}, R., \& {Zinger},
  E. 2009, \nat, 457, 451

\bibitem[{{Dijkstra} {et~al.}(2006){Dijkstra}, {Haiman}, \&
  {Spaans}}]{dijkstra2006a}
{Dijkstra}, M., {Haiman}, Z., \& {Spaans}, M. 2006, \apj, 649, 14

\bibitem[{{Dijkstra} {et~al.}(2007){Dijkstra}, {Lidz}, \&
  {Wyithe}}]{dijkstra2007}
{Dijkstra}, M., {Lidz}, A., \& {Wyithe}, J.~S.~B. 2007, \mnras, 377, 1175

\bibitem[{{Erb} {et~al.}(2003){Erb}, {Shapley}, {Steidel}, {Pettini},
  {Adelberger}, {Hunt}, {Moorwood}, \& {Cuby}}]{erb2003}
{Erb}, D.~K., {Shapley}, A.~E., {Steidel}, C.~C., {Pettini}, M., {Adelberger},
  K.~L., {Hunt}, M.~P., {Moorwood}, A.~F.~M., \& {Cuby}, J. 2003, \apj, 591,
  101

\bibitem[{{Erb} {et~al.}(2006{\natexlab{a}}){Erb}, {Steidel}, {Shapley},
  {Pettini}, {Reddy}, \& {Adelberger}}]{erb2006c}
{Erb}, D.~K., {Steidel}, C.~C., {Shapley}, A.~E., {Pettini}, M., {Reddy},
  N.~A., \& {Adelberger}, K.~L. 2006{\natexlab{a}}, \apj, 647, 128

\bibitem[{{Erb} {et~al.}(2006{\natexlab{b}}){Erb}, {Steidel}, {Shapley},
  {Pettini}, {Reddy}, \& {Adelberger}}]{erb2006b}
---. 2006{\natexlab{b}}, \apj, 646, 107

\bibitem[{{Fardal} {et~al.}(2001){Fardal}, {Katz}, {Gardner}, {Hernquist},
  {Weinberg}, \& {Dav{\'e}}}]{fardal01}
{Fardal}, M.~A., {Katz}, N., {Gardner}, J.~P., {Hernquist}, L., {Weinberg},
  D.~H., \& {Dav{\'e}}, R. 2001, \apj, 562, 605

\bibitem[{{Fumagalli} {et~al.}(2011){Fumagalli}, {Prochaska}, {Kasen}, {Dekel},
  {Ceverino}, \& {Primack}}]{fumagalli2011}
{Fumagalli}, M., {Prochaska}, J.~X., {Kasen}, D., {Dekel}, A., {Ceverino}, D.,
  \& {Primack}, J.~R. 2011, ArXiv e-prints

\bibitem[{{Hansen} \& {Oh}(2006)}]{hansenoh06}
{Hansen}, M. \& {Oh}, S.~P. 2006, \mnras, 367, 979

\bibitem[{{Kelson}(2003)}]{kelson2003}
{Kelson}, D.~D. 2003, \pasp, 115, 688

\bibitem[{{Kennicutt}(1998{\natexlab{a}})}]{kennicuttAR98}
{Kennicutt}, Jr., R.~C. 1998{\natexlab{a}}, \araa, 36, 189

\bibitem[{{Kennicutt}(1998{\natexlab{b}})}]{kennicutt98}
---. 1998{\natexlab{b}}, \apj, 498, 541

\bibitem[{{Kere{\v s}} {et~al.}(2009){Kere{\v s}}, {Katz}, {Fardal},
  {Dav{\'e}}, \& {Weinberg}}]{keres2009}
{Kere{\v s}}, D., {Katz}, N., {Fardal}, M., {Dav{\'e}}, R., \& {Weinberg},
  D.~H. 2009, \mnras, 395, 160

\bibitem[{{Kere{\v s}} {et~al.}(2005){Kere{\v s}}, {Katz}, {Weinberg}, \&
  {Dav{\'e}}}]{keres2005}
{Kere{\v s}}, D., {Katz}, N., {Weinberg}, D.~H., \& {Dav{\'e}}, R. 2005,
  \mnras, 363, 2

\bibitem[{{Kollmeier} {et~al.}(2010){Kollmeier}, {Zheng}, {Dav{\'e}}, {Gould},
  {Katz}, {Miralda-Escud{\'e}}, \& {Weinberg}}]{kollmeier2010}
{Kollmeier}, J.~A., {Zheng}, Z., {Dav{\'e}}, R., {Gould}, A., {Katz}, N.,
  {Miralda-Escud{\'e}}, J., \& {Weinberg}, D.~H. 2010, \apj, 708, 1048

\bibitem[{{Kornei} {et~al.}(2010){Kornei}, {Shapley}, {Erb}, {Steidel},
  {Reddy}, {Pettini}, \& {Bogosavljevi{\'c}}}]{kornei10}
{Kornei}, K.~A., {Shapley}, A.~E., {Erb}, D.~K., {Steidel}, C.~C., {Reddy},
  N.~A., {Pettini}, M., \& {Bogosavljevi{\'c}}, M. 2010, \apj, 711, 693

\bibitem[{{Laursen} {et~al.}(2009{\natexlab{a}}){Laursen}, {Razoumov}, \&
  {Sommer-Larsen}}]{laursen2009a}
{Laursen}, P., {Razoumov}, A.~O., \& {Sommer-Larsen}, J. 2009{\natexlab{a}},
  \apj, 696, 853

\bibitem[{{Laursen} {et~al.}(2009{\natexlab{b}}){Laursen}, {Sommer-Larsen}, \&
  {Andersen}}]{laursen2009b}
{Laursen}, P., {Sommer-Larsen}, J., \& {Andersen}, A.~C. 2009{\natexlab{b}},
  \apj, 704, 1640

\bibitem[{{Laursen} {et~al.}(2011){Laursen}, {Sommer-Larsen}, \&
  {Razoumov}}]{laursen2011}
{Laursen}, P., {Sommer-Larsen}, J., \& {Razoumov}, A.~O. 2011, \apj, 728, 52

\bibitem[{{Liu} {et~al.}(2008){Liu}, {Shapley}, {Coil}, {Brinchmann}, \&
  {Ma}}]{liu2008}
{Liu}, X., {Shapley}, A.~E., {Coil}, A.~L., {Brinchmann}, J., \& {Ma}, C. 2008,
  \apj, 678, 758

\bibitem[{{Madau}(1995)}]{madau95}
{Madau}, P. 1995, \apj, 441, 18

\bibitem[{{McLean} {et~al.}(1998){McLean}, {Becklin}, {Bendiksen}, {Brims},
  {Canfield}, {Figer}, {Graham}, {Hare}, {Lacayanga}, {Larkin}, {Larson},
  {Levenson}, {Magnone}, {Teplitz}, \& {Wong}}]{mclean1998}
{McLean}, I.~S., {Becklin}, E.~E., {Bendiksen}, O., {Brims}, G., {Canfield},
  J., {Figer}, D.~F., {Graham}, J.~R., {Hare}, J., {Lacayanga}, F., {Larkin},
  J.~E., {Larson}, S.~B., {Levenson}, N., {Magnone}, N., {Teplitz}, H., \&
  {Wong}, W. 1998, in Society of Photo-Optical Instrumentation Engineers (SPIE)
  Conference Series, Vol. 3354, Society of Photo-Optical Instrumentation
  Engineers (SPIE) Conference Series, ed. {A.~M.~Fowler}, 566--578

\bibitem[{{McLinden} {et~al.}(2011){McLinden}, {Finkelstein}, {Rhoads},
  {Malhotra}, {Hibon}, {Richardson}, {Cresci}, {Quirrenbach}, {Pasquali},
  {Bian}, {Fan}, \& {Woodward}}]{mclinden2011}
{McLinden}, E.~M., {Finkelstein}, S.~L., {Rhoads}, J.~E., {Malhotra}, S.,
  {Hibon}, P., {Richardson}, M.~L.~A., {Cresci}, G., {Quirrenbach}, A.,
  {Pasquali}, A., {Bian}, F., {Fan}, X., \& {Woodward}, C.~E. 2011, \apj, 730,
  136

\bibitem[{{Oke} {et~al.}(1995){Oke}, {Cohen}, {Carr}, {Cromer}, {Dingizian},
  {Harris}, {Labrecque}, {Lucinio}, {Schaal}, {Epps}, \& {Miller}}]{oke95}
{Oke}, J.~B., {Cohen}, J.~G., {Carr}, M., {Cromer}, J., {Dingizian}, A.,
  {Harris}, F.~H., {Labrecque}, S., {Lucinio}, R., {Schaal}, W., {Epps}, H., \&
  {Miller}, J. 1995, \pasp, 107, 375

\bibitem[{{Osterbrock}(1962)}]{osterbrock1962}
{Osterbrock}, D.~E. 1962, \apj, 135, 195

\bibitem[{{Osterbrock}(1989)}]{osterbrock89}
---. 1989, {Astrophysics of gaseous nebulae and active galactic nuclei}

\bibitem[{{Pettini} {et~al.}(2001){Pettini}, {Shapley}, {Steidel}, {Cuby},
  {Dickinson}, {Moorwood}, {Adelberger}, \& {Giavalisco}}]{pettini01}
{Pettini}, M., {Shapley}, A.~E., {Steidel}, C.~C., {Cuby}, J., {Dickinson}, M.,
  {Moorwood}, A.~F.~M., {Adelberger}, K.~L., \& {Giavalisco}, M. 2001, \apj,
  554, 981

\bibitem[{{Quider} {et~al.}(2009){Quider}, {Pettini}, {Shapley}, \&
  {Steidel}}]{quider09}
{Quider}, A.~M., {Pettini}, M., {Shapley}, A.~E., \& {Steidel}, C.~C. 2009,
  \mnras, 398, 1263

\bibitem[{{Rauch} {et~al.}(2011){Rauch}, {Becker}, {Haehnelt}, {Gauthier},
  {Ravindranath}, \& {Sargent}}]{rauch11}
{Rauch}, M., {Becker}, G.~D., {Haehnelt}, M.~G., {Gauthier}, J.-R.,
  {Ravindranath}, S., \& {Sargent}, W.~L.~W. 2011, ArXiv e-prints

\bibitem[{{Reddy} {et~al.}(2008){Reddy}, {Steidel}, {Pettini}, {Adelberger},
  {Shapley}, {Erb}, \& {Dickinson}}]{reddy08}
{Reddy}, N.~A., {Steidel}, C.~C., {Pettini}, M., {Adelberger}, K.~L.,
  {Shapley}, A.~E., {Erb}, D.~K., \& {Dickinson}, M. 2008, \apjs, 175, 48

\bibitem[{{Schaerer} \& {Verhamme}(2008)}]{schaerer2008}
{Schaerer}, D. \& {Verhamme}, A. 2008, \aap, 480, 369

\bibitem[{{Shapley} {et~al.}(2005){Shapley}, {Coil}, {Ma}, \&
  {Bundy}}]{shapley2005}
{Shapley}, A.~E., {Coil}, A.~L., {Ma}, C., \& {Bundy}, K. 2005, \apj, 635, 1006

\bibitem[{{Shapley} {et~al.}(2003){Shapley}, {Steidel}, {Pettini}, \&
  {Adelberger}}]{shapley03}
{Shapley}, A.~E., {Steidel}, C.~C., {Pettini}, M., \& {Adelberger}, K.~L. 2003,
  \apj, 588, 65

\bibitem[{{Steidel} {et~al.}(2003){Steidel}, {Adelberger}, {Shapley},
  {Pettini}, {Dickinson}, \& {Giavalisco}}]{steidel2003}
{Steidel}, C.~C., {Adelberger}, K.~L., {Shapley}, A.~E., {Pettini}, M.,
  {Dickinson}, M., \& {Giavalisco}, M. 2003, \apj, 592, 728

\bibitem[{{Steidel} {et~al.}(2011){Steidel}, {Bogosavljevi{\'c}}, {Shapley},
  {Kollmeier}, {Reddy}, {Erb}, \& {Pettini}}]{steidel11}
{Steidel}, C.~C., {Bogosavljevi{\'c}}, M., {Shapley}, A.~E., {Kollmeier},
  J.~A., {Reddy}, N.~A., {Erb}, D.~K., \& {Pettini}, M. 2011, ArXiv e-prints

\bibitem[{{Steidel} {et~al.}(2010){Steidel}, {Erb}, {Shapley}, {Pettini},
  {Reddy}, {Bogosavljevi{\'c}}, {Rudie}, \& {Rakic}}]{steidel10}
{Steidel}, C.~C., {Erb}, D.~K., {Shapley}, A.~E., {Pettini}, M., {Reddy}, N.,
  {Bogosavljevi{\'c}}, M., {Rudie}, G.~C., \& {Rakic}, O. 2010, \apj, 717, 289

\bibitem[{{Steidel} {et~al.}(2004){Steidel}, {Shapley}, {Pettini},
  {Adelberger}, {Erb}, {Reddy}, \& {Hunt}}]{steidel2004}
{Steidel}, C.~C., {Shapley}, A.~E., {Pettini}, M., {Adelberger}, K.~L., {Erb},
  D.~K., {Reddy}, N.~A., \& {Hunt}, M.~P. 2004, \apj, 604, 534

\bibitem[{{Tapken} {et~al.}(2007){Tapken}, {Appenzeller}, {Noll}, {Richling},
  {Heidt}, {Meink{\"o}hn}, \& {Mehlert}}]{tapken07}
{Tapken}, C., {Appenzeller}, I., {Noll}, S., {Richling}, S., {Heidt}, J.,
  {Meink{\"o}hn}, E., \& {Mehlert}, D. 2007, \aap, 467, 63

\bibitem[{{Veilleux} {et~al.}(2005){Veilleux}, {Cecil}, \&
  {Bland-Hawthorn}}]{veilleux05}
{Veilleux}, S., {Cecil}, G., \& {Bland-Hawthorn}, J. 2005, \araa, 43, 769

\bibitem[{{Verhamme} {et~al.}(2008){Verhamme}, {Schaerer}, {Atek}, \&
  {Tapken}}]{verhamme08}
{Verhamme}, A., {Schaerer}, D., {Atek}, H., \& {Tapken}, C. 2008, \aap, 491, 89

\bibitem[{{Verhamme} {et~al.}(2006){Verhamme}, {Schaerer}, \&
  {Maselli}}]{verhamme06}
{Verhamme}, A., {Schaerer}, D., \& {Maselli}, A. 2006, \aap, 460, 397

\bibitem[{{Yang} {et~al.}(2011){Yang}, {Zabludoff}, {Jahnke}, {Eisenstein},
  {Dav{\'e}}, {Shectman}, \& {Kelson}}]{yang2011}
{Yang}, Y., {Zabludoff}, A., {Jahnke}, K., {Eisenstein}, D., {Dav{\'e}}, R.,
  {Shectman}, S.~A., \& {Kelson}, D.~D. 2011, \apj, 735, 87

\bibitem[{{Zheng} {et~al.}(2010){Zheng}, {Cen}, {Trac}, \&
  {Miralda-Escud{\'e}}}]{zheng10}
{Zheng}, Z., {Cen}, R., {Trac}, H., \& {Miralda-Escud{\'e}}, J. 2010, \apj,
  716, 574

\bibitem[{{Zheng} \& {Miralda-Escud{\'e}}(2002)}]{zheng02}
{Zheng}, Z. \& {Miralda-Escud{\'e}}, J. 2002, \apj, 578, 33

\end{thebibliography}

\end{document}